\documentclass[useAMS]{mn2e}

\usepackage{graphicx}

\title[Chemical compositions of eclipsing close binaries]
{Compositional differences between the component stars of eclipsing 
close binary systems showing chemical peculiarities
}

\author[Y. Takeda et al.]
{Yoichi Takeda$^{1,2}$\thanks{E-mail:
takeda.yoichi@nao.ac.jp}\footnotemark[0], 
Inwoo Han$^{3,4}$, 
Dong-Il Kang$^{5}$,
Byeong-Cheol Lee$^{3,4}$, 
Kang-Min Kim$^{3,4}$
\\
$^{1}$National Astronomical Observatory of Japan, 2-21-1 Osawa, Mitaka, Tokyo 181-8588, Japan\\
$^{2}$SOKENDAI, The Graduate University for Advanced Studies, 2-21-1 Osawa, Mitaka, Tokyo 181-8588, Japan\\
$^{3}$Korea Astronomy and Space Science Institute, 776, Daedeokdae-ro, Youseong-gu, Daejeon 34055, Korea\\
$^{4}$Korea University of Science and Technology, 217, Gajeong-ro Youseong-gu, Daejeon 34113, Korea\\
$^{5}$Changwon Science high school, 30, Pyungsanro 159-th, Uichang, Changwon, 641-500, Korea
}
\begin{document}

\date{Accepted 2019 February 9. Received 2019 February 8; in original form 2018 December 3.}


\maketitle

\label{firstpage}

\begin{abstract}
A spectroscopic study was carried out on the surface chemical abundances of CNO and
several heavier elements in the primary and secondary components of 5 eclipsing 
close binaries around A-type (AR~Aur, $\beta$~Aur, YZ~Cas, WW~Aur, and RR~Lyn), 
in order to investigate the nature of chemical differences between both components
(being comparatively slow rotators alike due to tidal synchronization).
Regarding the systems comprising similar components, $\beta$~Aur and WW~Aur were
confirmed to exhibit no compositional difference between the primary 
and secondary both showing almost the same Am anomaly,
though the chemical peculiarities seen in the component stars 
of AR~Aur show distinct differences (HgMn star and Am star). In contrast, as to
the systems (YZ~Cas and RR~Lyn) consisting of considerably different (A and early-F) 
components, the surface abundances are markedly different between the primary (Am) 
and secondary (normal). These observational results may indicate 
$T_{\rm eff}$-dependent characteristics regarding the chemical anomalies of 
non-magnetic stars on the upper main sequence:
(1) In the effective temperature range of $10000 {\rm K} \ga T_{\rm eff} \ga 7000 {\rm K}$, 
rotational velocity is the most important factor for determining the extent 
of Am peculiarity. 
(2) However, the emergence of Am phenomenon seems to 
have a lower $T_{\rm eff}$ limit at $\sim 7000$~K, below which no abundance anomaly 
is observed regardless of stellar rotation. 
(3) The transition from Am 
anomaly (mild deficiency in CNO) to HgMn anomaly (unusually large N depletion) 
is likely to take place as $T_{\rm eff}$ increases from $\sim$~10000~K to $\sim$~11000~K.
\end{abstract}


\section{Introduction}
 
The nature and origin of chemically peculiar stars (CP stars) seen in late B 
through early F type stars of upper main sequence, which are divided into
several subgroups according to the type of peculiarity (e.g., Preston 1974), 
has long been a matter of debate. Among these, commonly observed CP phenomena 
among non-magnetic stars\footnote{This term ``non-magnetic stars'' is used to discriminate 
them from those CP stars showing strong magnetic fields (up to $\sim 10^{3}$--10$^{4}$ 
gauss) as well as specific abundance anomalies (e.g., overabundances of 
Cr, Sr, or Eu). Recent high-precision polarimetric observations revealed that 
weak magnetic fields do exist even in Am or HgMn stars.} are HgMn stars
(mostly seen in late B dwarfs) and Am stars (metallic-lined A stars, 
which are widely seen in early-A through early-F main-sequence stars).
Takeda et al. (2018c; hereinafter referred to as Paper~I) very recently carried 
out an extensive study of C, N, and O abundances for 100 sample stars (comprising 
normal stars as well as Am/HgMn stars), motivated by the situation that 
systematic investigation on these important light elements had been lacking,
and arrived at the following results:\\ 
--- (1) Almost all stars (irrespective of being classified as peculiar or normal, 
though with a tendency of larger anomaly for the former) show more or less CNO 
deficits typically in the range of $-1 \la $~[C,N,O/H]~$\la 0$; i.e., no stars 
turned out to show normal CNO abundances. Besides, distinctly large deficiencies 
as much as $\sim 2$~dex are also shown for C or N by some CP stars ([C/H] for 
late Am stars or [N/H] for HgMn stars of late B-type).\\
--- (2) More useful information could be extracted by concentrating on the 
homogeneous sample of 16 A-type stars belonging to the Hyades cluster, which must 
have the same age and the same initial composition. In this way, we could confirm 
the systematic increase of abundance peculiarity with a decrease in $v_{\rm e}\sin i$ 
(projected rotational velocity), which manifestly indicates that rotation is 
the key factor responsible for controlling the extent of Am anomaly.

Therefore, result~(1) suggests that CNO abundances are useful indicators for
judging the degree of chemical peculiarity, which are applicable to wide variety of 
late B through early F stars regardless of whether or not being classified as peculiar.
Besides, we learn from result~(2) that studying and comparing the abundances of stars 
sharing homogeneous character (e.g., born at the same time with the same composition) 
is quite an effective approach for clarifying the nature of chemical anomaly.  

It thus occurred to us to pay attention to the behavior of CNO abundances   
in the components of double-line spectroscopic binaries, which must have 
formed concurrently from the same gas such as the case of cluster stars. 
Distinct merits are also accompanied especially with eclipsing close binary systems, 
because stellar parameters (mass $M$, radius $R$, orbital inclination $i_{\rm orb}$, etc) 
are well defined. In addition, since similar rotational velocities are realized for
both components due to tidal synchronization, we would be able to check
whether any other factor (apart from stellar rotation) affecting the chemical    
anomaly exists.
Unfortunately, it is not an easy task to establish the abundances for both 
components of a spectroscopic binary, because lines of two stars are overlapped  
in a complex time-variable way depending on the orbital phase. Actually, such 
abundance studies of direct analyzing double-line spectra, which have occasionally 
been carried out since old days, are not regarded as sufficiently reliable,
especially when weak lines are to be measured in a crowded spectrum.

However, the situation has dramatically changed nowadays thanks to 
progress of spectrum disentangling\footnote{
Some authors prefer to use the term ``disentangling'' to specifically 
imply iterative determination of orbital elements, which requires reliable 
radial velocities based on appropriately isolated spectra of both 
components, while simply using ``separation'' if only spectrum 
reconstruction of each component is involved. Note, however, that 
we use ``disentangling'' in the latter meaning in this paper.} 
 technique (e.g., Iliji\'{c} 2004 and 
the references therein), by which the genuine spectra of primary/secondary 
can be separately extracted from a set of double-line spectra sufficiently 
covering various orbital phases, making precise abundance determinations
feasible for both components. Nevertheless, past applications of this technique
to studying compositional differences of spectroscopic binaries comprising 
CP stars of late B through early F-type under question (Am stars, HgMn stars) 
are still only a few in number to our knowledge; e.g., Folsom et al. (2010) 
for AR~Aur (B9V+B9.6V), Pavlovski et al. (2014) for YZ~Cas (A2IV+F2V),
and Torres et al. (2015) for V501~Mon (A6m+F0); and besides these studies
do not seem to have paid much attention to CNO abundances of our interest.

Motivated by this situation, we decided to investigate the abundances of 
CNO (as well as several other elements derived as by-products) and how they 
differ between the components for 5 eclipsing binary systems comprising 
CP stars (AR~Aur, $\beta$~Aur, YZ~Cas, WW~Aur, and RR~Lyn), based on 
the adequately disentangled spectra of the primary and secondary, 
in essentially the same manner as done in Paper~I.
These 5 systems, which were selected because sufficient observational data 
covering various phases (necessary for spectrum disentangling) are available,
can be divided into two groups: ``similar-component'' group
(AR~Aur with late-B stars, $\beta$~Aur with early-A stars, and WW~Aur
with mid-A stars) and ``different-component'' group (YZ Cas with 
early-A and early-F stars, RR~Lyn with late-A and early-F stars).
In what follows are the check points we would like to elucidate in this study.
\begin{itemize}
\item
All the component stars of our targets (with spectral types of late B 
through early F) are slow rotators ($v_{\rm e} \sim$~10--40~km~s$^{-1}$) 
because rotation and orbital motion are almost synchronized 
in such close binary systems (orbital period is $P \sim$~3--10~day). 
Then, according to our knowledge, we may expect that all these stars 
would more or less exhibit chemical peculiarities. Could we really 
confirm this?
\item
Regarding the systems of ``similar-component'' group, almost the same 
abundances should be observed for both components according to our naive 
expectation because differences in stelar parameters (including rotational
velocities) are fairly small. What about the observational results 
on this point?
\item
As to the binary systems belonging to ``different-component'' group, 
the primary and secondary stars have distinctly different $T_{\rm eff}$
from each other. Could we detect any compositional differences
between the two? If so, this may provide us with useful information
regarding $T_{\rm eff}$-dependence of chemical peculiarities. 
\end{itemize}

\section{Observational data}

\subsection{Observations}

The observations of our 5 program stars (AR~Aur, $\beta$~Aur, YZ~Cas, 
WW~Aur, and RR~Lyn; cf. Table~1 for the basic data) were carried out 
on 2010 December 14, 15, 16, 18, and 20 by using BOES 
(Bohyunsan Observatory Echelle Spectrograph) attached to the 1.8 m reflector 
at Bohyunsan Optical Astronomy Observatory. Using 2k$\times$4k CCD 
(pixel size of 15~$\mu$m~$\times$~15~$\mu$m), this echelle spectrograph 
enabled us to obtain spectra of wide wavelength coverage 
(from $\sim$~3800~\AA\ to $\sim$~9200~\AA).
We used 200$\mu$m fiber corresponding to the resolving power 
of $R \simeq 45000$. The total integration time for one observation 
(consisting of 2--3 successive frames to be co-added) was chosen to be 
$\sim$~10--60~minutes depending on the brightness of a target.
Each of the program stars were observed 1--3 times in a night with an interval 
of a few hours, though the actual frequency differed from star to star.
Thus, as a result of 5-night observations, we could obtain 53 spectra for 
these 5 targets, which consist of 9--11 spectra per each star corresponding 
to different observational times. The fundamental information 
(spectrum code and observational time in Julian day) for each of 
the 53 spectra is presented in Table 2.

The reduction of the echelle spectra (bias subtraction, flat 
fielding, spectrum extraction, wavelength calibration, and 
continuum normalization) was carried out by using the software package 
IRAF (Image Reduction and Analysis Facility).\footnote{
IRAF is distributed by the National Optical Astronomy Observatories,
which is operated by the Association of Universities for Research
in Astronomy, Inc. under cooperative agreement with the National 
Science Foundation.}

\subsection{Spectrum disentangling}

It is essential to correctly specify the local radial velocities for 
the primary ($V_{1}^{\rm local}$) and the secondary ($V_{2}^{\rm local}$) 
corresponding to each spectrum, in order to successfully obtain the 
disentangled spectra of both components. For this purpose, we first 
evaluated the predicted $V_{1}^{\rm local}$ and $V_{2}^{\rm local}$ at 
each relevant observational time as presented in Table~2 (7th and 8th 
columns), where we employed the ephemeris data of primary minimum (taken 
from the web site of General Catalogue of Variable Stars\footnote{Available 
at http://www.sai.msu.su/gcvs/gcvs/index.htm.}; Samus et al. 2017) 
and the orbital elements (taken from the compilation of Eker et al. 2014) 
summarized in Table~1. Then, we simulated the theoretical double-line 
profile of O~{\sc i} 7771--5 triplet corresponding to such evaluated  
$V_{1}^{\rm local}$ and $V_{2}^{\rm local}$ by using the spectrum-synthesis 
technique, which was further compared with the observed profile in order 
to check whether both agree well with each other.
While a satisfactory consistency was confirmed for $\beta$~Aur, YZ~Cas,
and WW~Aur, small systematic differences (due to slight variation of 
system velocity?) were seen for AR~Aur and appreciable phase-dependent 
disagreements (likely to be caused by inadequately computed phases) were 
observed for RR~Lyn.\footnote{
Although we refrain from discussing the reason for the discrepancy, 
possible anomalies of radial velocity curves have occasionally been 
mentioned for these two stars. See, e.g., Wilson \& Van Hamme (2014; Sect.~8 
and the references therein) for AR~Aur and Tomkin \& Fekel (2006; Sect.~6.1 
and the references therein) for RR~Lyn.} 
Therefore, we estimated the empirical radial velocities ($V_{1}^{\rm emp}$ 
and  $V_{2}^{\rm emp}$) by directly comparing the theoretical O~{\sc i} 7771--5 
line profiles (simulated for various combinations of $V_{1}$ and $V_{2}$)
with the observed one. 
These various kinds of radial velocities evaluated for each of the spectra 
are summarized in Table~2. 
The radial velocities of both components finally adopted for disentangling 
($V_{1}^{\rm local}$ and $V_{2}^{\rm local}$ for $\beta$~Aur, YZ~Cas,
and WW~Aur; $V_{1}^{\rm emp}$ and $V_{2}^{\rm emp}$ for AR~Aur and RR~Lyn) 
are graphically depicted by symbols (converted to the heliocentric scale) 
in Fig.~1, where the predicted radial velocity curves are also shown
for comparison. Note that 2 spectra for WW~Aur (20101216A) and RR~Lyn (20101214A) 
corresponding to eclipse phases were eventually not used, as indicated in 
Fig.~1 (by arrows) as well as in Table~2 (by parentheses). 

We selected four wavelength regions of 50--60~\AA\ width, for which 
disentangled spectra of both components are to be extracted: 5350--5400~\AA, 
6130--6180~\AA, 7440--7500~\AA, and 7750--7800~\AA.
The first three are obvious because we intend to determine the abundances of
C, N, and O based on C~{\sc i} 5380, N~{\sc i} 7468, and O~{\sc i} 6156--8 lines,
closely following Paper~I.  The inclusion of 7750--7800~\AA\ region is to make 
use of the O~{\sc i}~7771--5 triplet lines, which are so strong and 
saturated as to be applied to checking $v_{\rm t}$ (microturbulence) and to 
precisely determining $v_{\rm e}\sin i$ (cf. Sect.~4.1).
Another parameter necessary to specify for spectrum disentangling is   
the luminosity ratio of the primary and secondary ($l_{1} : l_{2}$)
relevant for each wavelength region. We computed this ratio by using the
emergent continuum fluxes ($F_{1}$ and $F_{2}$) computed from Kurucz's (1993) 
ATLAS9 program (corresponding to $T_{\rm eff}$ and $\log g$  for each star)
and the stellar radii ($R_{1}$ and $R_{2}$) as 
$l_{1} : l_{2} = F_{1}R_{1}^{2} : F_{2}R_{2}^{2}$, where the stellar parameters
were taken from Table~1. The resulting luminosity ratios are summarized in Table~3.

Regarding the software tool for spectrum disentangling, we made use of the 
public-domain program CRES\footnote{http://sail.zpf.fer.hr/cres/} 
written by S. Iliji\'{c}, as employed by Takeda, Hashimoto \& Honda (2018b)
for their study of Capella's photospheric abundances. See Sect.~3 of their paper 
for more details of the procedures we adopted. We confirmed that the resulting 
disentangled spectra turned out quite satisfactory. The accomplished S/N ratios 
are sufficiently high ($\sim$~300--700), except for RR~Lyn~(2) ($\sim$~100--200) 
and YZ~Cas~(2) ($\sim$~70--90), as summarized in Table~4. Fig.~2 demonstrates 
how the pure primary and secondary spectra could be successfully extracted 
from a set of original spectra in the neighborhood of O~{\sc i} 7771--5. 
All these disentangled spectra (for 10 stars and 4 regions) used in this study
are presented as on-line material (file name: all\_spectra).

Finally, it should be remarked that our application of the spectrum disentangling 
method implicitly assumed that the line spectra and the brightness of each 
component did not show any time variability during the observed period,
despite that the existence of small intrinsic spectrum variability 
(caused by chemical inhomogeneity on the surface?) has been reported 
for some of our program stars; e.g., for WW~Aur (cf. Kitamura, Kim \& Kiyokawa 1976)
and AR~Aur (cf. Folsom et al. 2010 and the references therein).   

\section{Abundance determination}

\subsection{Atmospheric parameters} 

Regarding $T_{\rm eff}$ and $\log g$ necessary to specify the model atmosphere
for each star, we adopted the values compiled by Eker et al. (2014) in
this study (cf. Table~1). Here, we should note that, while these $\log g$ values 
are regarded as fairly precise (quoted errors are only 0.01--0.03~dex) for these 
eclipsing binary stars because $M$ and $R$ are well established, $T_{\rm eff}$'s 
are the photometry-based values and may not be so accurate. 
In Fig.~3 are plotted our 10 program stars on the $T_{\rm eff}$ vs. $\log g$ plane, 
where theoretical evolutionary tracks of solar-metallicity stars computed for 
various $M$ values are also depicted. We can see from this figure that the  
track-based $M$ is not consistent with the real $M$ for some cases (e.g.,
for WW~Aur or RR~Lyn), for which $T_{\rm eff}$ errors even larger than the nominal 
values ($\sim$~100--400~K) may be possible. The microturbulent velocities 
($v_{\rm t}$) were derived from $T_{\rm eff}$ by using the empirical relation 
[Eq.~(1) in Takeda et al. (2008)] as done in Paper~I, though appreciable ambiguities
may be involved with them (cf. Sect.~4.1). In estimating abundance errors
due to uncertainties in atmospheric parameters, we tentatively assume in this study  
$\pm 500$~K in $T_{\rm eff}$, $\pm 0.1$~dex in $\log g$, and $\pm 50$\% in $v_{\rm t}$. 
for all stars (which may be regarded rather as upper limits of actual errors).

\subsection{Procedures}

The determination of CNO abundances (synthetic spectrum fitting, evaluation of 
equivalent widths, error estimation) was done in essentially the same manner as 
adopted in Paper~I, which should be consulted for the details (cf. Sect.~4 therein).
We here remark only some differences specific to this study.
\begin{itemize}
\item
The O~{\sc i} 7771--5 triplet lines are also analyzed based on the 7750--7800~\AA\ 
fitting, in a similar manner to the case of O~{\sc i} 6156--8, though this is  
mainly for the derivation of $v_{\rm e}\sin i$ as well as for checking $v_{\rm t}$
(rather than for O-abundance determination).
\item
The wavelength range of spectrum fitting targeting O~{\sc i} 6156--8 lines  
was extended somewhat shortwards to include Ba~{\sc ii} 6141 line, in order to  
get information of Ba abundance.
\item
We make use of the abundances of various elements other than CNO
(specifically, Na, Si, Ca, Ti, Fe, and Ba), which are derived as by-products 
of spectrum fitting (though under the assumption of LTE), since they also serve 
as useful indicators of chemical peculiarities. 
\end{itemize}

The details of spectrum fitting analysis (wavelength range, source of line data, 
varied abundances) done for four regions are presented in Table~5, and the 
basic atomic data of C~{\sc i} 5380, N~{\sc i} 7468, O~{\sc i} 6156--8,
and O~{\sc i} 7771--5 lines are summarized in Table~6.
The accomplished consistency between the theoretical spectrum (for the converged 
solutions) and the observed spectrum as a result of our fitting analysis is 
displayed in Fig.~4--7 for each region. In Fig.~ 8 are plotted the resulting
equivalent width ($W$), non-LTE correction ($\Delta$), and non-LTE abundance 
($A^{\rm N}$) for each of these CNO lines against $T_{\rm eff}$, just like Fig.~5--7
of Paper~I. 
More complete results of these CNO abundances (including errors and related 
quantities) and the abundances of other elements (Na, Si, Ca, Ti, Fe, and Ba) 
resulting from spectrum fitting as by-products are presented as online
material (filename: abundances.dat).  
Besides, the abundances of Na, Si, Ca, and Ti for 100 A-type main-sequence 
stars targeted in Paper~I, which were not published therein because the main 
focus was placed only on CNO, are also given as online material (filename: 
paperI\_suppl.dat) for reference.

\subsection{Impact of errors in the luminosity ratio}

Here, as an additional error source specific to this study, it may be worth mentioning 
the effect of uncertainties in  the luminosity ratio adopted for spectrum 
disentangling (cf. Sect.~2.2), since the emergent flux ($F$) would suffer some 
ambiguities due to errors in  $T_{\rm eff}$ (typically by a few hundred K; 
cf. Sect.~3.1), although the radii ($R$) of both stars are precisely determined.
Since $F$ is changed by $\sim$~20--30\% for an increase in $T_{\rm eff}$ by 1000~K,
errors of several to $\la 10$\% may be possible in the computed luminosity.  
As a test trial, we carried out spectrum disentangling in the 6130--6180~\AA\ region 
by increasing the $l_{1}/l_{2}$ ratio by 10\% (e.g., 0.542:0.458 instead of 
0.518:0.482 for AR~Aur or 0.793:0.207 instead of 0.777:0.223 for RR~Lyn),
and determined the abundances of O, Na, Si, Ca, Fe, and Ba in the same manner 
based on such specially reconstructed spectra with intentionally increased $l_{1}/l_{2}$. 
Comparing the resulting abundances with the standard values, we found that the
differences are generally insignificant (typically several hundredths dex),
except for the strong/deep line cases of Ba or Ca (cf. Fig~6) where more appreciable 
abundance changes of $\sim$~0.1--0.15~dex are seen (especially for YZ~Cas (2) and RR~Lyn (2)). 
This may be interpreted as due to the fact that an error in the zero-point level of 
the residual intensity, which is caused by inappropriately chosen luminosity ratio, 
significantly affects deep/strong lines but not shallow/weak lines (see footnote~12
in Takeda, Hashimoto, \& Honda 2018b). Accordingly, as far as the abundances of
C, N, and O derived from C~{\sc i} 5380, N~{\sc i} 7468, and O~{\sc i} 6156--8
lines (our main objective which we will discuss in Sect.~4.2) are concerned, this 
kind of error is of negligible importance, because these lines are sufficiently weak.

\section{Discussion}

\subsection{Microturbulence and rotation}

We start by mentioning about the stellar parameters ($v_{\rm t}$ and $v_{\rm e}\sin i$) 
related to O~{\sc i} 7771--5 triplet, which was specially analyzed in this study. 

Regarding the microturbulence, we simply applied the empirical $T_{\rm eff}$-dependent 
analytical formula proposed by Takeda et al. (2008), as done in Paper~I.
A useful test for examining the adequacy of such assumed $v_{\rm t}$ is to check 
whether the non-LTE oxygen abundances derived from strong O~{\sc i} 7771--5 lines 
($A^{\rm N}_{77}$) and weak O~{\sc i} 6156--8 lines ($A^{\rm N}_{61}$) agree with each other, 
because the former is appreciably $v_{\rm t}$-sensitive while the latter is not.    
Such a comparison is displayed in Fig.~9a, indicating that $A^{\rm N}_{77}$ 
tends to be somewhat larger than $A^{\rm N}_{61}$ (this difference is also  
recognized by comparing Fig.~8c and Fig.~8d). We can determine the
``abundance-matched'' microturbulence $v_{\rm t}$(A) by requiring 
the abundance equality ($A^{\rm N}_{77} = A^{\rm N}_{61}$; cf. Fig~9b). 
These $v_{\rm t}$(A) values are plotted against $T_{\rm eff}$ in Fig.~9c,
which shows that these values are somewhat larger than the analytical 
relation we adopted (as expected). For reference, we also derived the
``profile-based'' microturbulence $v_{\rm t}$(P), which was so determined
as to accomplish the best-fit O~{\sc i} 7771--5 triplet line profile 
(cf. Sect.~5 in Takeda, Jeong \& Han 2018a).
Such obtained $v_{\rm t}$(P) results are depicted in Fig.~9d, where we see
that $v_{\rm t}$(P) tends to become higher than the adopted relation at 
$T_{\rm eff} \la 8000$~K. Taking all these results into consideration,
we conservatively assumed that uncertainties of $\pm 50$\% may be involved with 
the $v_{\rm t}$ values adopted in this study, as mentioned in Sect.~3.1.   
Fortunately, the CNO abundances derived from C~{\sc i} 5380, N~{\sc i} 7468,
and O~{\sc i} 6156--8 lines (which are all quite weak) are hardly affected 
by any choice of $v_{\rm t}$, though abundances from stronger saturated lines 
(e.g., Ba abundance from Ba~{\sc ii} 6141 line) are considerably $v_{\rm t}$-dependent. 

Since the O~{\sc i} 7771--5 triplet is almost free from blending with other
lines and shows a characteristic profile comprising three component lines 
(which sometimes show a complex substructure by merging of rotationally 
broadened profiles; see, e.g., Fig.~7b or Fig.~7d), precise determination 
of $v_{\rm e}\sin i$ may be expected if successful fitting
between theoretical and observed profiles could be accomplished.
Such spectroscopically established $v_{\rm e}\sin i$ values (presented in 
Table~1) can be used to check whether the expected synchronization of rotational 
and orbital motions are actually realized in these close binary systems.
Assuming the alignment of orbital and rotational axes, we can derive the observed 
equatorial rotation velocity ($v_{\rm e}$) by dividing $v_{\rm e} \sin i$ by the sinus 
of orbital inclination ($i_{\rm orb}$). The comparison of such obtained $v_{\rm e}$ 
with $V_{\rm syn}$ ($\equiv 2\pi R/P$: synchronized rotation velocity) is shown 
in Fig.~10a, where can see that the agreement is mostly satisfactory. 
This suggests that synchronization is almost realized in these close binary systems,
which lends support to the result of Kitamura \& Kondo (1978) who arrived at 
practically the same conclusion. 
However, the situation is indefinite regarding the comparatively longer period case 
of RR~Lyn ($P \simeq 10$~day), where $v_{\rm e}$ is apparently larger than $V_{\rm syn}$ 
by $\sim$~20--40\% (cf. Fig.~10b). This might indicate that co-rotation has not yet 
been realized in this system. Alternatively, we can not rule out a possibility of 
overestimated $v_{\rm e}\sin i$ for such a sharp-lined star ($\sim 10$~km~s$^{-1}$) 
due to our neglect of some significant broadening component (e.g., macroturbulence), 
although the effect of instrumental broadening (corresponding to the spectrum resolving 
power of $R \simeq 45000$) was taken into consideration in deriving $v_{\rm e}\sin i$.

\subsection{Abundance characteristics}

We now discuss the abundances of 10 stars (primary and secondary of 5 systems) 
resulting from our analysis. Regarding oxygen, we exclusively use the abundance derived from 
O~{\sc i} 6156--8 lines (as done in Paper~I), because the abundance from 
O~{\sc i} 7771--5 is sensitive to $v_{\rm t}$ and suffers a large non-LTE effect.
The differential abundances relative to the standard star Procyon\footnote{
The surface composition of this star is known to be almost the same as that of the Sun.
See the references quoted in Sect.~IV(c) of Takeda et al. (2008).} ([X/H]; X is any of 
C, N, O, Na, Si, Ca, Ti, Fe, and Ba) for each star are summarized in Table~7 and 
graphically plotted in Fig.~11. We can read from this figure several characteristic 
trends in terms of the check points set up in Sect.~1.\\
--- Regarding the systems belonging to ``similar component'' group (AR~Aur, $\beta$~Aur,
WW~Aur), the abundances of the primary and secondary components are very similar and
hardly discernible for $\beta$~Aur and WW~Aur, while those for AR~Aur are apparent different.
More precisely, both components of $\beta$~Aur and WW~Aur show the characteristics of 
Am phenomena: C/N/O are deficient by several tenths to $\la 1$~dex (cf. Paper~I),  
Si/Na/Ti/Fe are normal/enriched, Ca is normal/underabundant, and Ba is 
conspicuously overabundant (see, e.g., Fig.~9 in Varenne \& Monier 1999). 
In contrast, the primary component of AR~Aur shows a clear characteristics of HgMn stars
(this star is actually known to be of this type) in the sense that N is drastically 
depleted by $\ga 2$~dex, whereas the secondary appears to exhibit some weak Am phenomenon 
as seen from moderate deficiencies of CNO and large overabundance of Ba.\\
--- As for YZ~Cas and RR~Lyn belonging to ``different-component'' systems, the surface 
composition of the hotter primary is appreciably different from that of the cooler secondary.
That is, the former shows clear Am-type peculiarities mentioned above, while the latter 
is nearly normal within a few tenths dex without showing any conspicuous anomaly
despite of slow rotators. It is worth noting that these normal-composition secondary 
stars of YZ~Cas and RR~Lyn are at $T_{\rm eff} \la 7000$~K.

Combining these observational facts, we can draw the following consequences regarding 
the appearance and properties of chemical peculiarities in late B through early F stars.
\begin{itemize}
\item
The fact that essentially the same Am peculiarities are observed between the similar primary 
and secondary (quite similar $v_{\rm e}\sin i$ and only small $T_{\rm eff}$ difference 
of $\sim$~200--300~K) for both $\beta$~Aur and WW~Aur means that Am phenomenon is mainly 
determined by $v_{\rm e}\sin i$ and not very sensitive to $T_{\rm eff}$ as far as the
$T_{\rm eff}$ range of $10000 {\rm K} \ga T_{\rm eff} \ga 7000 {\rm K}$ is concerned. 
This is a reconfirmation of the primary importance of rotational velocity in the 
appearance of Am anomalies, though their extents show moderate $T_{\rm eff}$-dependences 
(e.g., tendency of larger CNO deficiency towards lower $T_{\rm eff}$).
\item
In the context of essentially same chemical compositions between the primary 
and secondary for $\beta$~Aur and WW~Aur, it is interesting to note that these two 
systems show markedly different Na as well as Ca abundances from each other (cf. 
Fig.~11b,d). We consider that this discrepancy is mainly due to the difference 
in $T_{\rm eff}$ (by $\sim 1500$~K), 
since [Na/H] and [Ca/H] tend to decrease with a lowering of $T_{\rm eff}$ 
as can be recognized in Fig.~12, where the [Na/H] and [Ca/H] values
of 100 A-type stars studied in Paper~I are plotted against $T_{\rm eff}$.
Likewise, diffusion calculation 
predicts the similar tendency at least for Ca (see, e.g., Fig.~14 in Richer, 
Michaud \& Turcotte 2000). Although these two stars have distinctly different ages,
$\sim$~450--500~Myr for $\beta$~Aur (Southworth, Bruntt \& Buzasi 2007) and 
$\sim 90$~Myr for WW~Aur (Southworth et al. 2005), it seems rather unlikely that 
this has caused the abundance difference, because Ca is considerably deficient in
younger WW~Aur while nearly normal in older $\beta$~Aur (i.e., in conflict with
the naive expectation that chemical anomaly would develop with time).   
\item
The apparent abundance difference between AR~Aur~(1) (HgMn star) and AR~Aur~(2) (weak Am star)
despite their similarity in stellar parameters (essentially the same rotational velocity 
and only a small $T_{\rm eff}$ difference of 600~K) is somewhat strange. It may be possible
that a rather distinct transition occurs somewhere in-between $T_{\rm eff} \sim$~10000--11000~K 
regarding the chemical peculiarity of slow rotators; i.e., from Am-type to HgMn-type 
as $T_{\rm eff}$ is increased. 
\item
The chemical composition difference between the primary (Am) and the secondary (normal) 
seen in YZ~Cas ($T_{\rm eff,1}/T_{\rm eff,2}$ = 9200~K/6700~K) and RR~Lyn (7570~K/6980~K) 
provides us with important information concerning the $T_{\rm eff}$-dependence
for the appearance of Am stars. Especially, the distinct transition from Am
to normal over a small $T_{\rm eff}$ difference of $\sim 600$~K for the case of RR~Lyn
is significant. We can state from these facts that the advent of Am phenomenon is 
confined to $T_{\rm eff} \ga 7000$~K, below which chemical peculiarities do not
appear no matter how slowly a star rotates. Actually, we can see from Fig.~8e--8g 
of Paper~I that the large spread of C, N, and O abundance deficiencies tend to 
quickly shrink at $T_{\rm eff} \sim 7000$~K. This is presumably related to the
thickening of surface convection zone counteracting the formation of chemical anomaly.  
\end{itemize}

Further, in order to visualize these abundance characteristics summarized above, 
we plot in Fig.~13 the values of [C/H], [N/H], and [O/H] for the 10 program stars 
as well as those of 100 stars (Hyades stars and field stars) taken from Paper~I 
altogether against $T_{\rm eff}$ and $v_{\rm e}\sin i$.
It can be recognized from this figure that the CNO abundances of the close 
binary stars are more or less consistent with the results for general 
A-type stars. Especially, our results (except for YZ~Cas (2) and RR~Lyn (2), 
which are normal abundance stars at $T_{\rm eff} \la 7000$~K) 
reasonably follow the systematic [X/H] vs. $v_{\rm e}\sin i$ trend established by 
Hyades stars (small open circles) despite the narrow range of $v_{\rm e}\sin i$ 
($\sim$~10--40~km~s$^{-1}$), which reaffirms the importance of rotational
velocity for the build-up of Am peculiarity.  

\subsection{Comparison with previous studies}

Finally, we briefly comment on how the results derived from our spectroscopic analysis 
of 5 binary systems are compared with similar chemical abundance studies published so far.
Here, focus of attention is placed on whether or not these previous investigations 
are consistent with our consequences. 

\subsubsection{AR~Aur}

Khokhlova et al. (1995) concluded in their chemical abundance study of AR~Aur
for 18 elements (based on the line strengths of each component carefully measured 
from the double-lined spectra of various phases) that the primary is a typical
HgMn star while the secondary is also abnormal but with different type of peculiarity.
Similarly, according to Folsom et al.'s (2010) more recent investigation based on 
the disentangled spectra of AR~Aur, the surface chemical compositions of the primary 
and secondary are apparently different, indicating that the former is a HgMn star
and the latter is a weak Am star (cf. Fig.~4 therein). Therefore, our conclusion is
almost consistent with these past studies. Although Folsom et al. (2010) reported
that the C and O abundances are nearly solar for both components (in contrast to
our result of mild but definite deficiency), this may be partly due to their neglect 
of the non-LTE effect.

\subsubsection{$\beta$~Aur}

Lyubimkov, Rachkovskaya \& Rostopchin (1996) determined the chemical compositions 
of the primary and secondary stars of $\beta$~Aur by analyzing the double-lined 
spectra, and concluded that the abundances of both components are quite similar and show 
the characteristics of Am stars. The abundance characteristics of various elements 
(including CNO) reported by them are in satisfactory agreement with our results.   

\subsubsection{YZ~Cas}

Pavlovski et al. (2014) recently carrird out chemical abundance determinations separately 
for each of the components of YZ~Cas based on the disentangled spectra. They reported 
that the primary is an Am star with supersolar abundances (except for Sc) while 
the composition of secondary is nearly solar, which is almost consistent with our conclusion. 
However, their results of ``supersolar'' C and O abundances (like other heavier elements)
for the Am primary are in serious conflict with the general trend of CNO deficiency 
confirmed in Paper~I as well as in this study. Unfortunately, since they did not 
publish any details of their analysis (e.g., used lines, equivalent widths), we have 
no idea about the reason for this discrepancy.

\subsubsection{WW~Aur}

To our knowledge, no chemical abundance study for this binary system has ever
been published so far. Although Pavlovski, Southworth \& Tamajo (2008) once reported 
their accomplishment of spectrum disentangling applied to WW~Aur, stating that the 
related abundance analysis was in progress, their successive paper does not seem to
have come out yet. 

\subsubsection{RR~Lyn}

Several spectroscopic investigations have already been published since old days
regarding the chemical abundances of RR~Lyn (Popper 1971; Kondo 1976; 
Burkhart \& Coupry 1991;  Lyubimkov \& Rachkovskaya 1995; Hui-Bon-Hoa 2000),
although all these are based on the direct analysis of double-lined spectra.
According to these previous studies, the primary component is an Am star 
and more metal-rich than the secondary, which is in reasonable consistency 
with our consequence. However, none of them mentioned about the CNO abundances
of this binary system. 

\section{Conclusion}

Spectroscopically investigating the chemical compositions separately for the primary 
and secondary components of a double-line eclipsing binary showing characteristic 
chemical peculiarities (typically seen in A-tye stars in the upper main sequence)
is of profound significance, since it may provide us with information regarding
how the appearance of chemical anomaly depends upon stellar physical parameters.

We conducted a spectroscopic study to determine the surface chemical abundances 
of CNO and other heavier elements in the primary and secondary components of selected 
5 eclipsing close binaries around A-type (AR~Aur, $\beta$~Aur, YZ Cas, WW~Aur, and RR~Lyn).
Our aim was to investigate whether any chemical difference exists between
two components, and how the situation depends upon stellar parameters. 

For this purpose, we extracted the individual spectra of the primary and secondary for 
each target in 4 wavelength regions by applying the spectrum disentangling method 
to a set of time-series spectra sufficiently covering the orbital phases, which 
were obtained by using the high-dispersion echelle spectrograph attached to 
the 1.8~m reflector of Bohyunsan Optical Astronomy Observatory.

The abundances of C, N, and O were determined by using C~{\sc i} 5380, N~{\sc i}~7468, 
and O~{\sc i} 6156--8 lines in essentially the same manner (spectrum-synthesis analysis 
by taking into account the non-LTE effect) as done in Paper~I. Besides, we could 
also derive the abundances of Na, Si, Ca, Ti, Fe, and Ba as by-products of 
spectrum fitting.

In addition, we also analyzed the O~{\sc i} 7771--5 triplet lines in the same way, 
with an aim to check the adopted $v_{\rm t}$ and to determine $v_{\rm e}\sin i$ for each star.
We could confirm that synchronization between rotation and orbital motion is 
accomplished for almost all systems, except for RR~Lyn for which the situation is unclear. 

Regarding the ``similar-component'' systems, $\beta$~Aur and WW~Aur were confirmed 
to exhibit the characteristics of Am anomaly (deficiency in CNO, overabundance
of heavy elements such as Ba) with essentially the same compositions 
for both the primary and secondary.

However, distinct compositional differences were found between the primary and 
secondary of AR~Aur, another ``similar-component'' system, in the sense that they 
show chemical anomalies of different types  (HgMn star and Am star, respectively). 

In contrast, as to YZ~Cas and RR~Lyn, both of which are the ``different-component'' 
systems consisting of A-type and early-F stars, the surface abundances turned out 
markedly different between the primary (Am star) and secondary (chemically normal). 

These observational results may indicate the $T_{\rm eff}$-dependent nature in the chemical 
anomalies of non-magnetic stars on the upper main sequence, as summarized below.\\
---(i) In the effective temperature range of $10000 {\rm K} \ga T_{\rm eff} \ga 7000 {\rm K}$, 
slow rotation is considered to be the key factor for the appearance of Am anomaly.
Thus, the components of close binaries (slow rotators due to tidal braking) 
naturally exhibit very similar Am anomalies under the same rotational velocities
($\beta$~Aur and WW~Aur).\\ 
---(ii) However, the emergence of Am peculiarity seems to have a distinct lower 
$T_{\rm eff}$ limit at $\sim 7000$~K, below which abundance anomaly can hardly be 
produced regardless of stellar rotation (secondary components of RR~Lyn and YZ~Cas).\\
---(iii) On the other hand, as seen from the abundance differences between the components
of AR~Aur, the transition from Am anomaly to HgMn anomaly is likely to take place 
(with an increase in $T_{\rm eff}$) somewhere between $\sim$~10000~K and $\sim$~11000~K.

\section*{Acknowledgments}
We express our cordial thanks to Dr. S. Iliji\'{c} for kindly making the useful 
spectrum disentangling software CRES open to the public.
Data reduction and analysis were in part carried out by using the common-use data 
analysis computer system at the Astronomy Data Center (ADC) of the National Astronomical 
Observatory of Japan.

\onecolumn

\setcounter{table}{0}
\begin{table*}
\begin{minipage}{180mm}
\small
\caption{Fundamental data of the program stars.}
\begin{center}
\begin{tabular}{cccccccl}\hline
       &       &    AR~Aur    &  $\beta$~Aur   &  YZ~Cas   &   WW~Aur  &  RR~Lyn   &  Remark  \\
\hline
\multicolumn{2}{c}{[Basic data]} & & & & & & \\
HD      &       &    34364     &   40183    &   4161    &   46052   &  44691    &          \\
HIP     &       &    24740     &   28360    &   3572    &   31173   &  30651    &          \\
$V$     &(mag)    &     6.14     &   1.89     &   5.65    &   5.82    &  5.55     &          \\
$\pi$   &(m.a.s.) &     8.19     &   40.21    &   10.75   &   12.41   &  13.34    &          \\
Sp      &       &  B9V+B9.6V   & A1Vm+A1Vm  & A2IV+F2V  &  A4m+A5m  & A6IV+F0V  &          \\
\hline
\multicolumn{2}{c}{[Orbital elements]} & & & & & & \\
JD(min~I)&(day)    &  2452501.392 & 2452500.573&2454509.297&2452501.814&2433153.862& from GCVS\\
$P$     &(day)    &     4.134651 &    3.960036&   4.467222& 2.52501936&  9.945079 & from GCVS\\
$K_{1}$ &(km~s$^{-1}$)   &       107.20 &      107.75&      73.35&    116.81 &    66.00  &          \\
$K_{2}$ &(km~s$^{-1}$)   &       115.90 &      111.25&     125.70&    126.49 &    84.00  &          \\
$\gamma$&(km~s$^{-1}$)   &        25.10 &      $-17.00$&       8.14&     $-8.47$ &   $-12.00$  &          \\
$e$     &       &        0.000 &       0.002&      0.000&     0.000 &    0.079  &          \\
$i_{\rm orb}$ &(deg)    &        88.52 &       76.91&      88.30&     87.55 &    87.45  &          \\
\hline
\multicolumn{2}{c}{[Stellar parameters]} & & & & & & \\
$M_{1}$ &($M_{\odot}$)  &        2.480 &       2.376&      2.310&     1.964 &    1.927  &          \\
$M_{2}$ &($M_{\odot}$)  &        2.294 &       2.291&      1.350&     1.814 &    1.507  &          \\
$R_{1}$ &($R_{\odot}$)  &        1.781 &       2.762&      2.530&     1.927 &    2.570  &          \\
$R_{2}$ &($R_{\odot}$)  &        1.816 &       2.568&      1.350&     1.841 &    1.590  &          \\
$T_{\rm eff,1}$   &(K)      &        10950 &        9350&       9200&      7960 &     7570  &          \\
$T_{\rm eff,2}$   &(K)      &        10350 &        9200&       6700&      7670 &     6980  &          \\
$\log g_{1}$    &(dex)    &        4.331 &       3.932&      3.995&     4.162 &    3.900  &          \\
$\log g_{2}$  &(dex)    &        4.280 &       3.979&      4.307&     4.167 &    4.214  &          \\
$v_{\rm t,1}$  &(km~s$^{-1}$)   &          1.0 &         2.9&        3.0&       4.0 &      3.8  &  $v_{\rm t}(T_{\rm eff})$ formula \\
$v_{\rm t,2}$  &(km~s$^{-1}$)   &          1.6 &         3.0&        2.6&       3.9 &      3.1  &  $v_{\rm t}(T_{\rm eff})$ formula \\
$v_{\rm e}\sin i_{1}$  &(km~s$^{-1}$)   &         23.2 &        35.9&       29.3&      36.2 &     15.2  & O~{\sc i} 7771--5 fitting\\
$v_{\rm e}\sin i_{2}$  &(km~s$^{-1}$)   &         22.6 &        34.1&       13.2&      38.8 &     11.6  & O~{\sc i} 7771--5 fitting\\
\hline
\end{tabular}
\end{center}
The data of JD(min~I) and $P$ were taken from the web site of General Catalogue of Variable 
Stars (Samus et al. 2017). 
The values of $v_{\rm t,1}$ and $v_{\rm t,2}$ (microturbulence) were estimated 
by using the $T_{\rm eff}$-dependent relation derived by Takeda et al. (2008) [cf. 
Eq.~(1) therein], while $v_{\rm e}\sin i_{1}$ and $v_{\rm e}\sin_{2}$
(projected rotational velocity) are the results of O~{\sc i} 7771--5 fitting.
All other data were adopted from the compilation of Ecker et al. (2014), which should be
consulted for more details (e.g., original references, involved errors, etc.).  
\end{minipage}
\end{table*}

\setcounter{table}{1}
\begin{table*}
\begin{minipage}{180mm}
\small
\caption{Observational dates and radial velocities.}
\begin{center}
\begin{tabular}{cccrrrrrrr}\hline
Spectrum label& JD(+2400000) & Phase & $V_{1}^{\rm hel}$ & $V_{2}^{\rm hel}$ & $\delta V^{\rm hel}$ 
& $V_{1}^{\rm local}$ & $V_{2}^{\rm local}$ & $V_{1}^{\rm emp}$ & $V_{2}^{\rm emp}$ \\
(1) & (2) & (3) & (4) & (5) & (6) & (7) & (8) & (9) & (10) \\
\hline
araur\_20101214A&55545.010&0.125&$-$50.6 &107.0 &$-$0.6 &$-$50.1 &107.6 &$-$45.1&112.6 \\
araur\_20101214B&55545.201&0.171&$-$69.2 &127.0 &$-$1.0 &$-$68.1 &128.0 &$-$63.1&133.0 \\
araur\_20101215A&55546.022&0.370&$-$53.2 &109.8 &$-$1.1 &$-$52.1 &110.9 &$-$47.1&115.9 \\
araur\_20101215B&55546.272&0.430&$-$20.5 &74.4 &$-$1.7 &$-$18.8 &76.1 &$-$13.8&81.1 \\
araur\_20101216A&55546.999&0.606&91.2 &$-$46.4 &$-$1.6 &92.8 &$-$44.8 &97.8&$-$39.8 \\
araur\_20101218A&55548.961&0.080&$-$26.8 &81.2 &$-$2.6 &$-$24.2 &83.8 &$-$19.2&88.8 \\
araur\_20101218B&55549.163&0.129&$-$52.7 &109.2 &$-$3.0 &$-$49.6 &112.2 &$-$44.6&117.2 \\
araur\_20101218C&55549.284&0.159&$-$64.9 &122.4 &$-$3.3 &$-$61.6 &125.7 &$-$56.6&130.7 \\
araur\_20101220A&55550.982&0.569&70.3 &$-$23.7 &$-$3.7 &73.9 &$-$20.1 &79.9&$-$14.1 \\
araur\_20101220B&55551.152&0.610&93.6 &$-$48.9 &$-$4.0 &97.6 &$-$44.9 &103.6&$-$38.9 \\
araur\_20101220C&55551.295&0.645&109.8 &$-$66.5 &$-$4.3 &114.1 &$-$62.1 &120.1&$-$56.1 \\
\hline
betaur\_20101214A&55545.043&0.798&86.1 &$-$122.9 &3.7 &82.4 &$-$126.6 &  \\
betaur\_20101214B&55545.255&0.852&69.6 &$-$106.0 &3.3 &66.4 &$-$109.3 &  \\
betaur\_20101215A&55546.087&0.062&$-$57.8 &25.2 &3.1 &$-$60.9 &22.1 &  \\
betaur\_20101215B&55546.261&0.106&$-$83.4 &51.7 &2.7 &$-$86.1 &49.0 &  \\
betaur\_20101216A&55547.023&0.298&$-$119.7 &89.3 &2.7 &$-$122.5 &86.6 &  \\
betaur\_20101218A&55548.984&0.793&87.0 &$-$123.9 &1.8 &85.2 &$-$125.7 &  \\
betaur\_20101218B&55549.129&0.830&77.6 &$-$114.3 &1.5 &76.1 &$-$115.8 &  \\
betaur\_20101218C&55549.314&0.877&58.4 &$-$94.6 &1.2 &57.3 &$-$95.8 &  \\
betaur\_20101220A&55551.005&0.304&$-$118.5 &88.0 &0.8 &$-$119.3 &87.3 &  \\
betaur\_20101220B&55551.203&0.354&$-$102.7 &71.3 &0.4 &$-$103.1 &70.9 &  \\
betaur\_20101220C&55551.356&0.393&$-$84.6 &52.3 &0.2 &$-$84.7 &52.2 &  \\
\hline
yzcas\_20101214A&55544.903&0.823&73.9 &$-$104.5 &$-$5.9 &79.8 &$-$98.5 &  \\
yzcas\_20101214B&55545.136&0.875&59.9 &$-$80.5 &$-$6.1 &66.0 &$-$74.4 &  \\
yzcas\_20101216A&55546.968&0.286&$-$63.4 &130.7 &$-$6.5 &$-$56.9 &137.2 &  \\
yzcas\_20101218A&55548.906&0.719&80.1 &$-$115.2 &$-$6.9 &87.0 &$-$108.4 &  \\
yzcas\_20101218B&55549.071&0.756&81.4 &$-$117.5 &$-$7.0 &88.5 &$-$110.4 &  \\
yzcas\_20101218C&55549.233&0.793&78.9 &$-$113.1 &$-$7.1 &85.9 &$-$106.0 &  \\
yzcas\_20101220A&55550.905&0.167&$-$55.4 &117.1 &$-$7.4 &$-$48.1 &124.5 &  \\
yzcas\_20101220B&55551.101&0.211&$-$63.0 &130.0 &$-$7.5 &$-$55.5 &137.5 &  \\
yzcas\_20101220C&55551.241&0.242&$-$65.1 &133.7 &$-$7.5 &$-$57.6 &141.2 &  \\
\hline
wwaur\_20101214A&55545.053&0.234&$-$124.7 &117.4 &7.5 &$-$132.3 &109.9 &  \\
wwaur\_20101214B&55545.262&0.317&$-$115.0 &106.9 &7.1 &$-$122.1 &99.9 &  \\
wwaur\_20101215A&55546.054&0.631&77.1 &$-$101.1 &7.0 &70.1 &$-$108.1 &  \\
wwaur\_20101215B&55546.204&0.690&100.2 &$-$126.2 &6.7 &93.5 &$-$132.8 &  \\
(wwaur\_20101216A)&55547.028&0.017&$-$20.7 &4.7 &6.6 &($-$27.2) &($-$1.8) &  \\
wwaur\_20101218A&55548.987&0.793&104.2 &$-$130.4 &5.6 &98.6 &$-$136.0 &  \\
wwaur\_20101218B&55549.134&0.851&85.8 &$-$110.5 &5.3 &80.5 &$-$115.8 &  \\
wwaur\_20101218C&55549.318&0.924&45.5 &$-$66.9 &4.9 &40.6 &$-$71.8 &  \\
wwaur\_20101220A&55551.009&0.593&56.1 &$-$78.4 &4.5 &51.6 &$-$83.0 &  \\
wwaur\_20101220B&55551.179&0.661&90.5 &$-$115.6 &4.2 &86.3 &$-$119.8 &  \\
wwaur\_20101220C&55551.326&0.719&106.1 &$-$132.6 &3.8 &102.3 &$-$136.4 &  \\
\hline
(rrlyn\_20101214A)&55545.084&0.487&$-$27.0 &$-$18.6 &5.2 &$-$32.2 &$-$23.8 &($-$5.0)&($-$30.0) \\
rrlyn\_20101214B&55545.292&0.508&$-$18.8 &$-$29.1 &4.9 &$-$23.7 &$-$33.9 &3.0&$-$40.0 \\
rrlyn\_20101215A&55546.163&0.596&17.7 &$-$65.5 &4.6 &13.1 &$-$70.1 &29.0&$-$74.0 \\
rrlyn\_20101215B&55546.303&0.610&23.4 &$-$69.9 &4.4 &19.0 &$-$74.3 &33.0&$-$77.0 \\
rrlyn\_20101216A&55547.051&0.685&48.9 &$-$86.1 &4.4 &44.5 &$-$90.4 &43.0&$-$90.0 \\
rrlyn\_20101218A&55549.010&0.882&35.9 &$-$64.3 &3.5 &32.3 &$-$67.8 &21.0&$-$62.0 \\
rrlyn\_20101218B&55549.193&0.901&29.1 &$-$57.7 &3.3 &25.8 &$-$61.0 &15.0&$-$53.0 \\
rrlyn\_20101218C&55549.368&0.918&22.2 &$-$50.8 &3.0 &19.2 &$-$53.8 &10.0&$-$45.0 \\
rrlyn\_20101220A&55551.032&0.086&$-$43.7 &33.5 &2.6 &$-$46.4 &30.9 &$-$55.0&39.0 \\
rrlyn\_20101220B&55551.206&0.103&$-$49.0 &42.1 &2.4 &$-$51.4 &39.8 &$-$62.0&47.0 \\
rrlyn\_20101220C&55551.361&0.119&$-$53.4 &49.4 &2.1 &$-$55.5 &47.2 &$-$67.0&54.0 \\
\hline
\end{tabular}
\end{center}
(1) Spectrum label (e.g., araur\_20101218C is the 3rd spectrum of AR~Aur observed on 
2010 December 18). (2) Heliocentric Julian day. (3) Orbital phase calculated from 
the $P$ and JD(min~I) values given in Table 1. (4) Predicted heliocentric 
radial velocity (in km~s$^{-1}$) for the primary calculated with the orbital elements 
given in Table~1. (5) Predicted heliocentric radial velocity for the secondary.
(6) Heliocentric correction (km~s$^{-1}$) computed by the ``rvcorrect'' task of IRAF. 
(7) Predicted local topocentric radial velocity (km~s$^{-1}$) for the primary.
(8) Predicted local topocentric radial velocity (km~s$^{-1}$) for the secondary.
(9) Empirically determined radial velocity of the primary directly measured from the spectrum
by comparing the simulated double-line profile of O~{\sc i} 7771--5 triplet with
the observed one.
(10) Empirically determined radial velocity of the secondary.
Note that $V_{1}^{\rm emp}$ and $V_{2}^{\rm emp}$ were used for spectrum
disentangling for AR~Aur and RR~Lyn, because predicted $V_{1}^{\rm local}$ and 
$V_{2}^{\rm local}$ did not satisfactorily reproduce the actual spectra for these two stars.
Otherwise, $V_{1}^{\rm local}$ and $V_{2}^{\rm local}$ were used for the other 3 stars,
which were confirmed to well match the observed spectra at each of the phases. 
Two spectra (wwaur\_20101216A and rrlyn\_20101214A, indicated by parentheses) were not 
used for disentangling, because they correspond to phases during the eclipse.

\end{minipage}
\end{table*}

\setcounter{table}{2}
\begin{table*}
\begin{minipage}{180mm}
\small
\caption{Adopted luminosity ratio $l_{1}:l_{2}$ at each wavelength region.}
\begin{center}
\begin{tabular}{cccccrcccccc}\hline
\hline
Star  &   5350--5400~\AA &   6130--6180~\AA &  7440--7500~\AA\ &  7750--7800~\AA\ \\
\hline
AR~Aur     & 0.520 : 0.480 & 0.518 : 0.482 & 0.515 : 0.485 & 0.515 : 0.485 \\
$\beta$~Aur& 0.545 : 0.455 & 0.544 : 0.456 & 0.543 : 0.457 & 0.543 : 0.457 \\
YZ~Cas     & 0.917 : 0.083 & 0.903 : 0.097 & 0.883 : 0.117 & 0.879 : 0.121 \\
WW~Aur     & 0.559 : 0.441 & 0.553 : 0.447 & 0.546 : 0.454 & 0.545 : 0.455 \\
RR~Lyn     & 0.788 : 0.212 & 0.777 : 0.223 & 0.764 : 0.236 & 0.761 : 0.239 \\
\hline
\end{tabular}
\end{center}
\end{minipage}
\end{table*}

\setcounter{table}{3}
\begin{table*}
\begin{minipage}{180mm}
\small
\caption{Accomplished S/N ratios of the disentangled spectra.}
\begin{center}
\begin{tabular}{cccccrcccccc}\hline
Star & $\sim 5380$~\AA & $\sim 6150$~\AA & $\sim 7460$~\AA & $\sim 7770$~\AA \\ 
\hline
AR~Aur~(1)  & 400 &  500 &  350  & 400 \\
AR~Aur~(2)  & 400 &  400 &  300  & 350 \\
$\beta$~Aur~(1) & 700 &  600 &  450  & 600 \\
$\beta$~Aur~(2) & 600 &  600 &  400  & 500 \\
YZ~Cas~(1)  & 600 &  700 &  500  & 600 \\
YZ~Cas~(2)  &  80 &   90 &   70  &  70 \\
WW~Aur~(1)  & 400 &  450 &  350  & 400 \\
WW~Aur~(2)  & 400 &  400 &  300  & 400 \\
RR~Lyn~(1)  & 500 &  600 &  350  & 600 \\
RR~Lyn~(2)  & 150 &  200 &  150  & 200 \\
\hline
\end{tabular}
\end{center}
\end{minipage}
\end{table*}

\setcounter{table}{4}
\begin{table*}
\begin{minipage}{180mm}
\small
\caption{Outline of spectrum-fitting analysis in this study.}
\begin{center}
\begin{tabular}{ccccc}\hline\hline
Main purpose & Fitting range (\AA) & Abundances varied$^{*}$ & Atomic data source & Figure \\
\hline
C abundance from C~{\sc i} 5380   & 5375--5390 & C, Ti, Fe & KB95m1 & Fig.~4 \\
N abundance from N~{\sc i} 7468  & 7457--7472 & N, Fe & KB95m2 & Fig.~5 \\
O abundance from O~{\sc i} 6156--8   & 6140--6168 & O, Na, Si, Ca, Fe, Ba & KB95 & Fig.~6 \\
$v_{\rm e}\sin i$ determination from O~{\sc i} 7771--5   & 7765--7785 & O, Fe & KB95m3 & Fig.~7 \\
\hline
\end{tabular}
\end{center}
$^{*}$ The abundances of all other elements than these were fixed in the fitting. \\
KB95m1 --- All the atomic line data presented in Kurucz \& Bell (1995) were used, 
excepting that the contribution of Fe~{\sc i} 5382.474 ($\chi_{\rm low} = 4.371$~eV) 
was neglected (because we found its $gf$ value to be erroneously too large). \\
KB95m2 --- All the atomic line data were taken from Kurucz \& Bell (1995), excepting
that the contribution of S~{\sc i} 7468.588 ($\chi_{\rm low} = 7.867$~eV) was neglected 
(because we found its $gf$ value to be erroneously too large). \\
KB95 --- All the atomic line data given in Kurucz \& Bell (1995) were used unchanged.\\
KB95m3 --- All the atomic line data were taken from Kurucz \& Bell (1995), except for
the $\log gf$ value of Fe~{\sc i} 7780.552 ($\chi_{\rm low} = 4.473$~eV), for which
we adopted $-0.066$ (cf. Takeda \& Sadakane 1997).
\end{minipage}
\end{table*}

\setcounter{table}{5}
\begin{table*}
\begin{minipage}{180mm}
\small
\caption{Adopted atomic data of relevant CNO lines.}
\begin{center}
\begin{tabular}{ccccccccc}\hline\hline
Line & Multiplet & Equivalent & $\lambda$ & $\chi_{\rm low}$ & $\log gf$ & Gammar & Gammas & Gammaw\\
     &  No.           &  Width  & (\AA) & (eV) & (dex) & (dex) & (dex) & (dex) \\  
\hline
C~{\sc i} 5380 & (11)  & $W_{5380}$ &  5380.337 & 7.685 & $-1.842$ & (7.89) & $-$4.66 & ($-$7.36)\\
\hline
N~{\sc i} 7468 & (3) & $W_{7468}$ & 7468.312 & 10.336 & $-0.270$  & 8.64 & $-$5.40 & ($-$7.60)\\
\hline
O~{\sc i} 6156--8 & (10) & $W_{6156-8}$ & 6155.961 & 10.740 & $-1.401$ & 7.60 & $-$3.96 & ($-$7.23)\\
(9 components)&  &  & 6155.971 & 10.740 & $-1.051$ & 7.61 & $-$3.96 & ($-$7.23)\\
         &         &          & 6155.989 & 10.740 & $-1.161$ & 7.61 & $-$3.96 & ($-$7.23)\\
         &         &          & 6156.737 & 10.740 & $-1.521$ & 7.61 & $-$3.96 & ($-$7.23)\\
         &         &          & 6156.755 & 10.740 & $-0.931$ & 7.61 & $-$3.96 & ($-$7.23)\\
         &         &          & 6156.778 & 10.740 & $-0.731$ & 7.62 & $-$3.96 & ($-$7.23)\\
         &         &          & 6158.149 & 10.741 & $-1.891$ & 7.62 & $-$3.96 & ($-$7.23)\\
         &         &          & 6158.172 & 10.741 & $-1.031$ & 7.62 & $-$3.96 & ($-$7.23)\\
         &         &          & 6158.187 & 10.741 & $-0.441$ & 7.61 & $-$3.96 & ($-$7.23)\\ 
\hline
O~{\sc i} 7771--5  & (1) &  $W_{7771-5}$ & 7771.944 & 9.146&+0.324 & 7.52 & $-$5.55 & ($-$7.65) \\
(3 components)     &     &               & 7774.166 & 9.146&+0.174 & 7.52 & $-$5.55 & ($-$7.65) \\
                   &     &               & 7775.388 & 9.146&$-$0.046 & 7.52 & $-$5.55 & ($-$7.65) \\
\hline
\end{tabular}
\end{center}
Following columns 3--5 (laboratory wavelength, lower excitation potential, 
and $\log gf$ value), three kinds of damping parameters are presented in columns 6--8:  
Gammar is the radiation damping width (s$^{-1}$) [$\log\gamma_{\rm rad}$], 
Gammas is the Stark damping width (s$^{-1}$) per electron density (cm$^{-3}$) 
at $10^{4}$ K [$\log(\gamma_{\rm e}/N_{\rm e})$], and
Gammaw is the van der Waals damping width (s$^{-1}$) per hydrogen density 
(cm$^{-3}$) at $10^{4}$ K [$\log(\gamma_{\rm w}/N_{\rm H})$]. \\
All the data were taken from Kurucz \& Bell (1995), except for 
the parenthesized damping parameters (unavailable in their compilation), 
for which the default values computed by Kurucz's (1993) WIDTH9 program were assigned.
\end{minipage}
\end{table*}

\setcounter{table}{6}
\begin{table*}
\begin{minipage}{180mm}
\small
\caption{Derived abundances of C, N, O, Na, Si, Ca, Ti, Fe, and Ba.}
\begin{center}
\begin{tabular}{cccccccccc}\hline
Star & [C] & [N] & [O] & [Na] & [Si] & [Ca] & [Ti] & [Fe] & [Ba] \\
(1)  & (2) & (3) & (4) & (5)  &  (6) &  (7) &  (8) &  (9) & (10) \\ 
\hline
AR~Aur~(1)      & $-$1.11:& $(<-1.9)$ & $-$0.43& +0.30& +0.25& +0.34& +0.51& +0.47& $-$0.05 \\
AR~Aur~(2)      & $-$0.68& $-$0.55& $-$0.31& $\cdots$ & $-$0.65& $-$0.24& $-$0.10& +0.07& +0.79 \\
$\beta$~Aur~(1) & $-$0.59& $-$0.68& $-$0.48& +0.44& +0.07& $-$0.01& +0.12& +0.32& +1.18 \\
$\beta$~Aur~(2) & $-$0.65& $-$0.71& $-$0.46& +0.43& +0.07& $-$0.06& +0.14& +0.37& +1.26 \\
YZ~Cas~(1)      & $-$0.59& $-$0.82& $-$0.43& +0.45& +0.16& $-$0.05& +0.23& +0.40& +1.43 \\
YZ~Cas~(2)      & $-$0.15& $-$0.11& $-$0.23& $-$0.21& $-$0.07& $-$0.23& $-$0.13& $-$0.19& $-$0.44 \\
WW~Aur~(1)      & $-$0.90& $-$1.00& $-$0.34& +0.04& $-$0.12& $-$1.05& $-$0.09& +0.09& +0.97 \\
WW~Aur~(2)      & $-$0.80& $-$0.85& $-$0.32& $-$0.02& $-$0.06& $-$1.08& $-$0.04& +0.15& +1.11 \\
RR~Lyn~(1)      & $-$0.99& $-$1.08& $-$0.72& +0.00& $-$0.12& $-$0.80& $-$0.22& +0.11& +1.21 \\
RR~Lyn~(2)      & $-$0.15& $-$0.20& +0.00& $-$0.38& $-$0.33& $-$0.30& $-$0.21& $-$0.41& $-$0.14 \\
\hline
\end{tabular}
\end{center}
All the results given here are the differential abundances relative to Procyon
(see ReadMe of the online material for the detailed reference abundances of Procyon determined 
in the same manner).\\ 
(2) Non-LTE C abundance from C~{\sc i} 5380. (3) Non-LTE N abundance from N~{\sc i} 7468.
(4) Non-LTE O abundance from O~{\sc i} 6156--8. (5) LTE Na abundance derived from
6140--6168~\AA\ fitting. (6) LTE Si abundance derived from 6140--6168~\AA\ fitting.
(7) LTE Ca abundance derived from 6140--6168~\AA\ fitting.
(8) LTE Ti abundance derived from 5375--5390~\AA\ fitting.
(9) Mean of the 4 LTE Fe abundances derived from the fittings of 5375--5390~\AA,
6140--6168~\AA, 7457--7472~\AA, and 7765--7785~\AA.
(10) LTE Ba abundance derived from 6140--6168~\AA\ fitting.\\
Regarding the data given for the primary of AR~Aur, the parenthesized value for N 
is the upper limit, while the value for C with colon (:) is considered to suffer 
a large uncertainty. 
\end{minipage}
\end{table*}

\clearpage

\setcounter{figure}{0}
\begin{figure*}
\begin{minipage}{140mm}
\includegraphics[width=14.0cm]{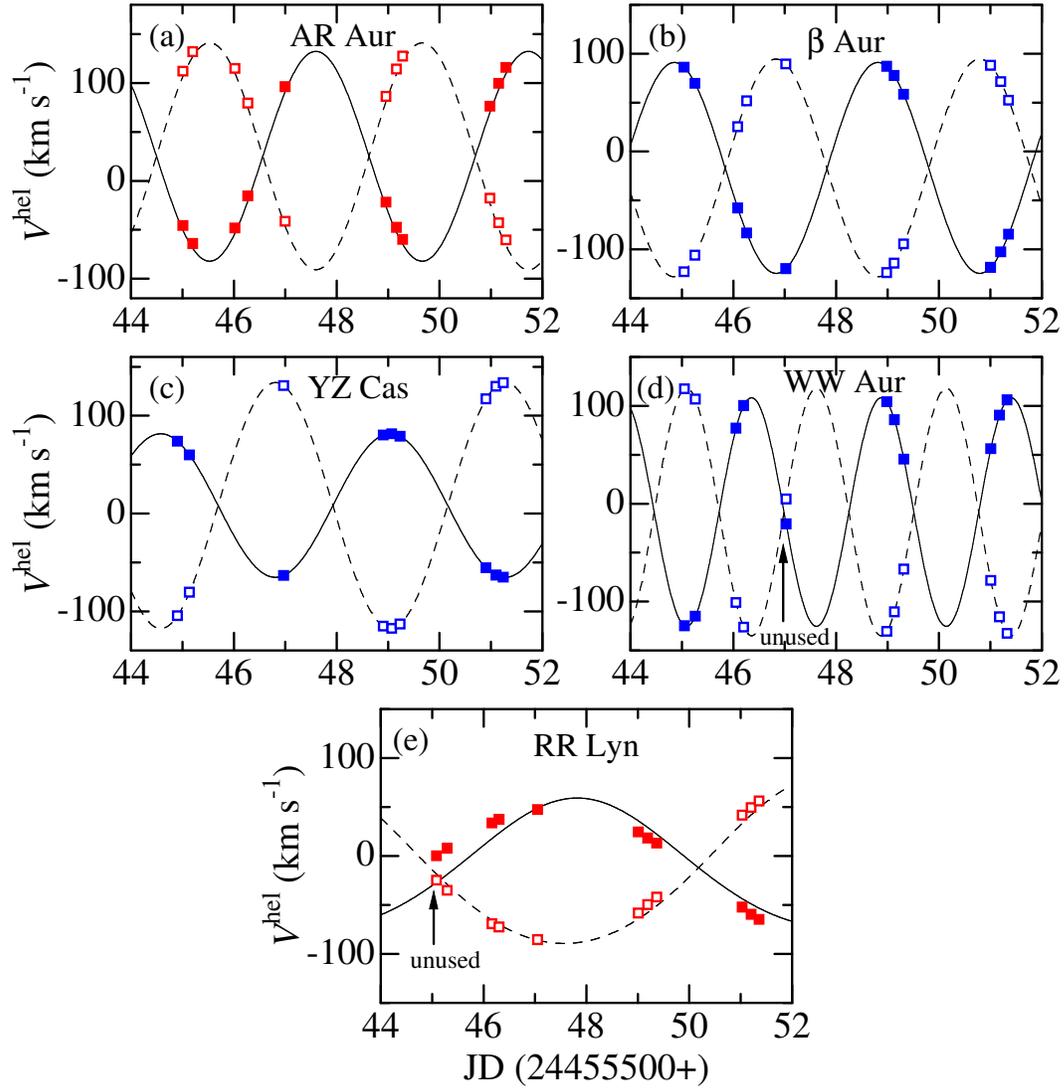}
\caption{
The solid (primary) and dashed (secondary) lines show the predicted heliocentric 
radial velocity curves calculated by using the orbital elements given in Table~1.
The filled (primary) and open (secondary) symbols indicate the radial velocities 
corresponding to each of the spectra, which were actually used for spectrum disentangling
(reduced to the heliocentric system). That is, according to the notation of Table~2, 
$V_{1}^{\rm local} + \delta V^{\rm hel}$ and $V_{2}^{\rm local} + \delta V^{\rm hel}$ 
(for $\beta$~Aur, YZ~Cas, and WW~Aur), or $V_{1}^{\rm emp} + \delta V^{\rm hel}$ 
and $V_{2}^{\rm emp} + \delta V^{\rm hel}$ (for AR~Aur and RR~Lyn). 
The data indicated by arrows in panels (d) and (e) were not used, because they
correspond to eclipse phases. (a) AR~Aur, (b) $\beta$~Aur, (c) YZ~Cas, 
(d) WW~Aur, and (e) RR~Lyn. 
}
\label{fig1}
\end{minipage}
\end{figure*}

\setcounter{figure}{1}
\begin{figure*}
\begin{minipage}{140mm}
\includegraphics[width=14.0cm]{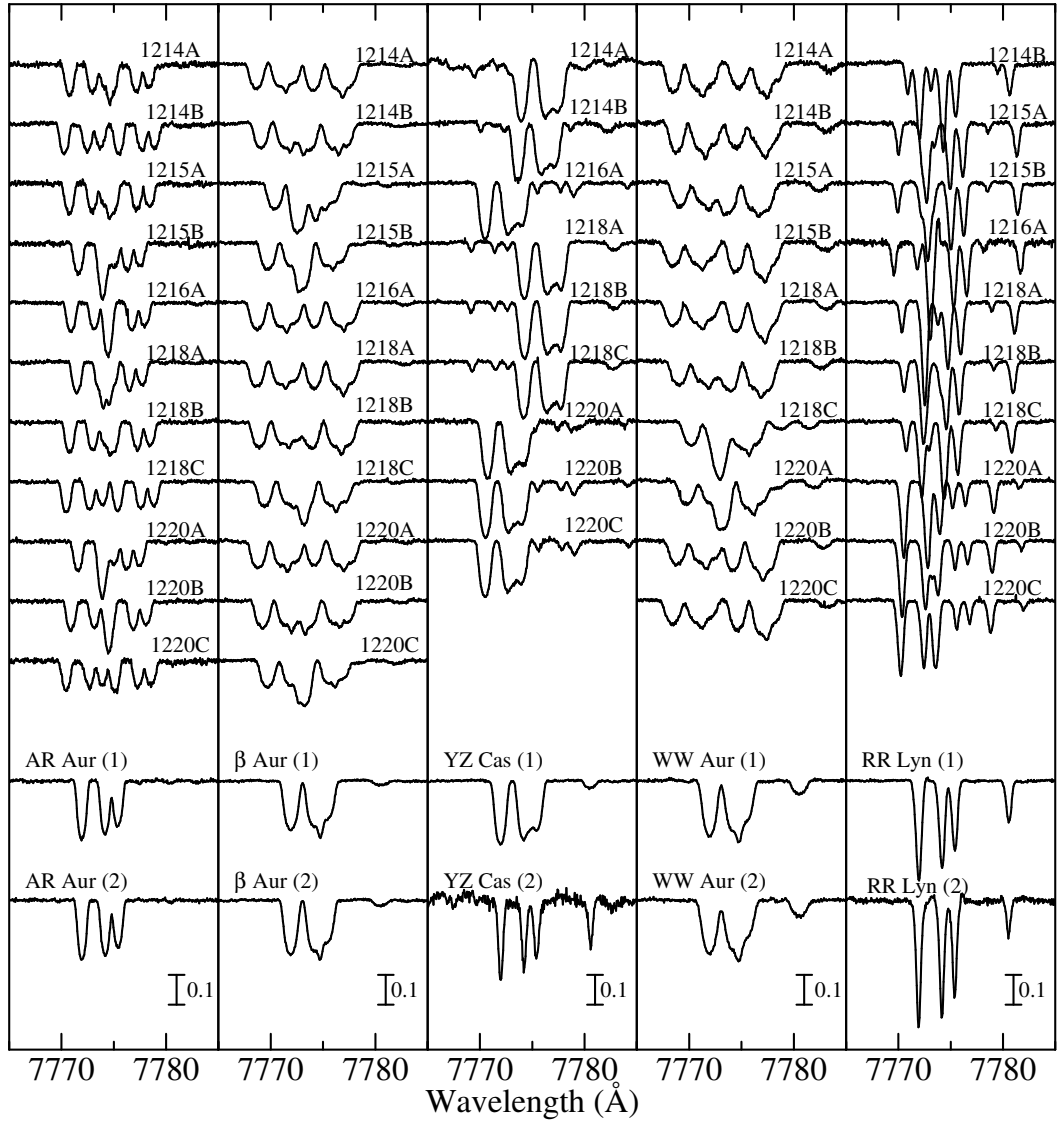}
\caption{
In each of the panels (from left to right: AR~Aur, $\beta$~Aur, YZ~Cas, WW~Aur, and RR~Lyn) 
are shown the original set of spectra used for spectrum entangling (indicated by the observed 
date and sequence) as well as the resulting disentangled spectra (the lowest two) in the 
neighborhood of the O~{\sc i} 7771--5 triplet lines. 
}
\label{fig2}
\end{minipage}
\end{figure*}

\setcounter{figure}{2}
\begin{figure*}
\begin{minipage}{80mm}
\includegraphics[width=8.0cm]{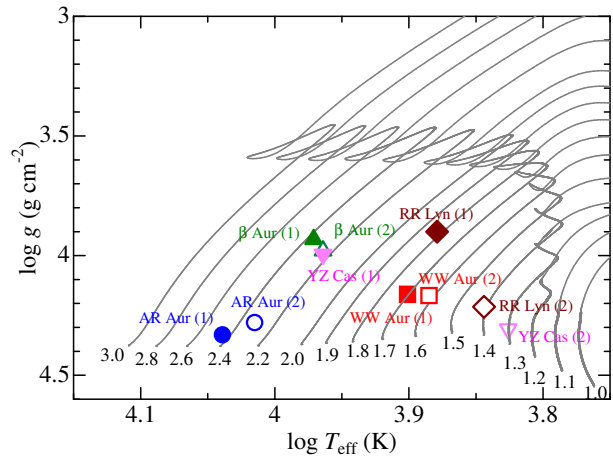}
\caption{
Our program stars plotted on the $T_{\rm eff}$ vs. $\log g$ diagram.
Each of the 5 binary systems are shown by different symbol shapes
[circles (AR~Aur), triangles ($\beta$~Aur), inverse triangles (YZ~Cas),
squares (WW~Aur), and diamonds (RR~Lyn)], while the filled and open
symbols correspond to the primary (1) and the secondary (2) components,
respectively. The solid lines show the theoretical relations predicted
from PARSEC stellar evolution calculations for solar-metallicity models 
(Bressan et al. 2012), where the corresponding stellar mass (in unit 
of $M_{\odot}$) is indicated for each track.
}
\label{fig3}
\end{minipage}
\end{figure*}

\setcounter{figure}{3}
\begin{figure*}
\begin{minipage}{140mm}
\includegraphics[width=14.0cm]{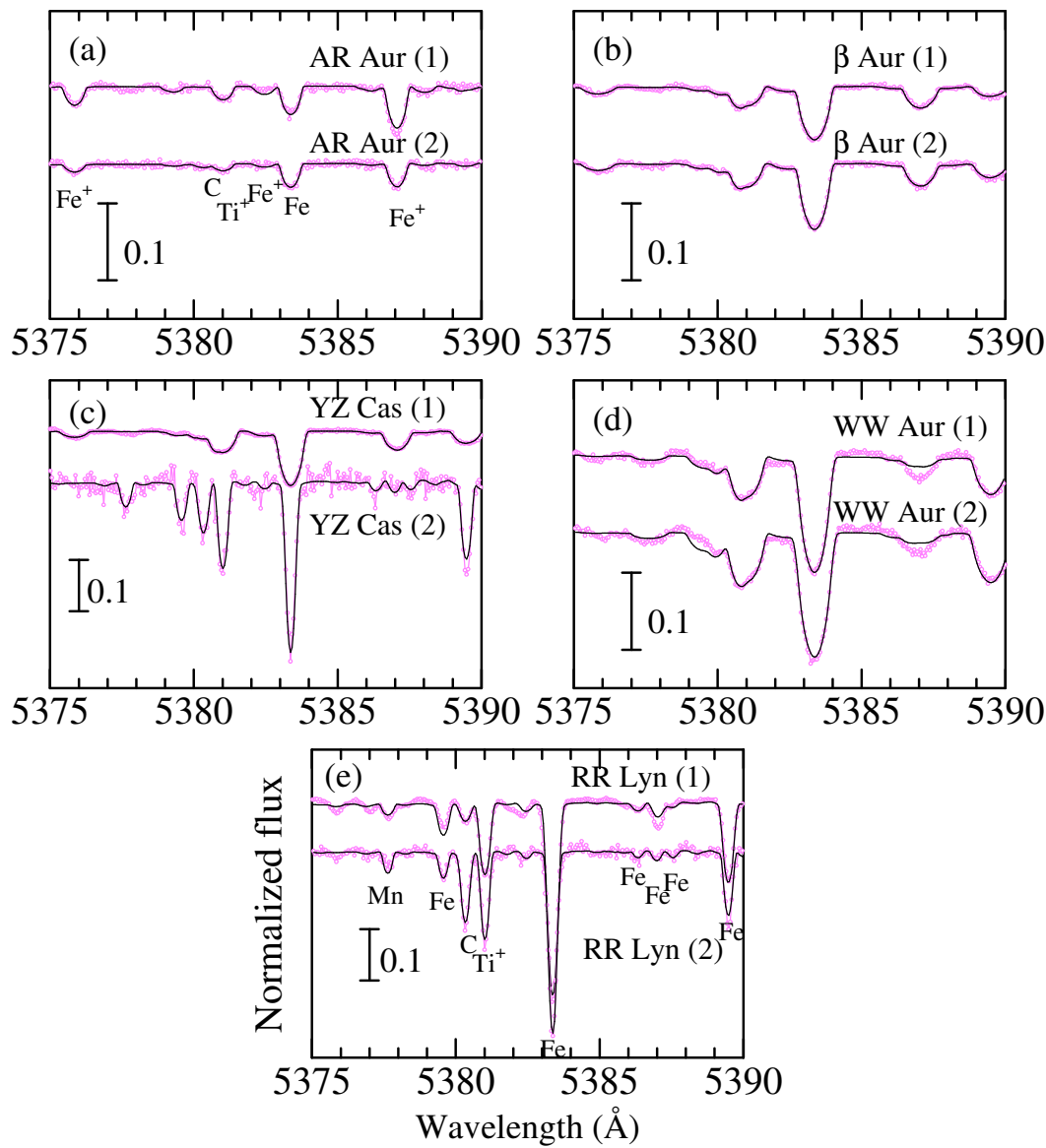}
\caption{
Synthetic spectrum fitting in the 5375--5390~\AA\ region 
comprising the C~{\sc i}~5380 line (along with the lines
of Ti and Fe). The best-fit theoretical spectra are shown by 
black solid lines. while the disentangled observed spectra are 
plotted by pink symbols. In each panel are shown the results 
for the primary (upper) as well as the secondary (lower). 
(a) AR~Aur, (b) $\beta$~Aur, (c) YZ~Cas, (d) WW~Aur, and (e) RR~Lyn. 
}
\label{fig4}
\end{minipage}
\end{figure*}

\setcounter{figure}{4}
\begin{figure*}
\begin{minipage}{140mm}
\includegraphics[width=14.0cm]{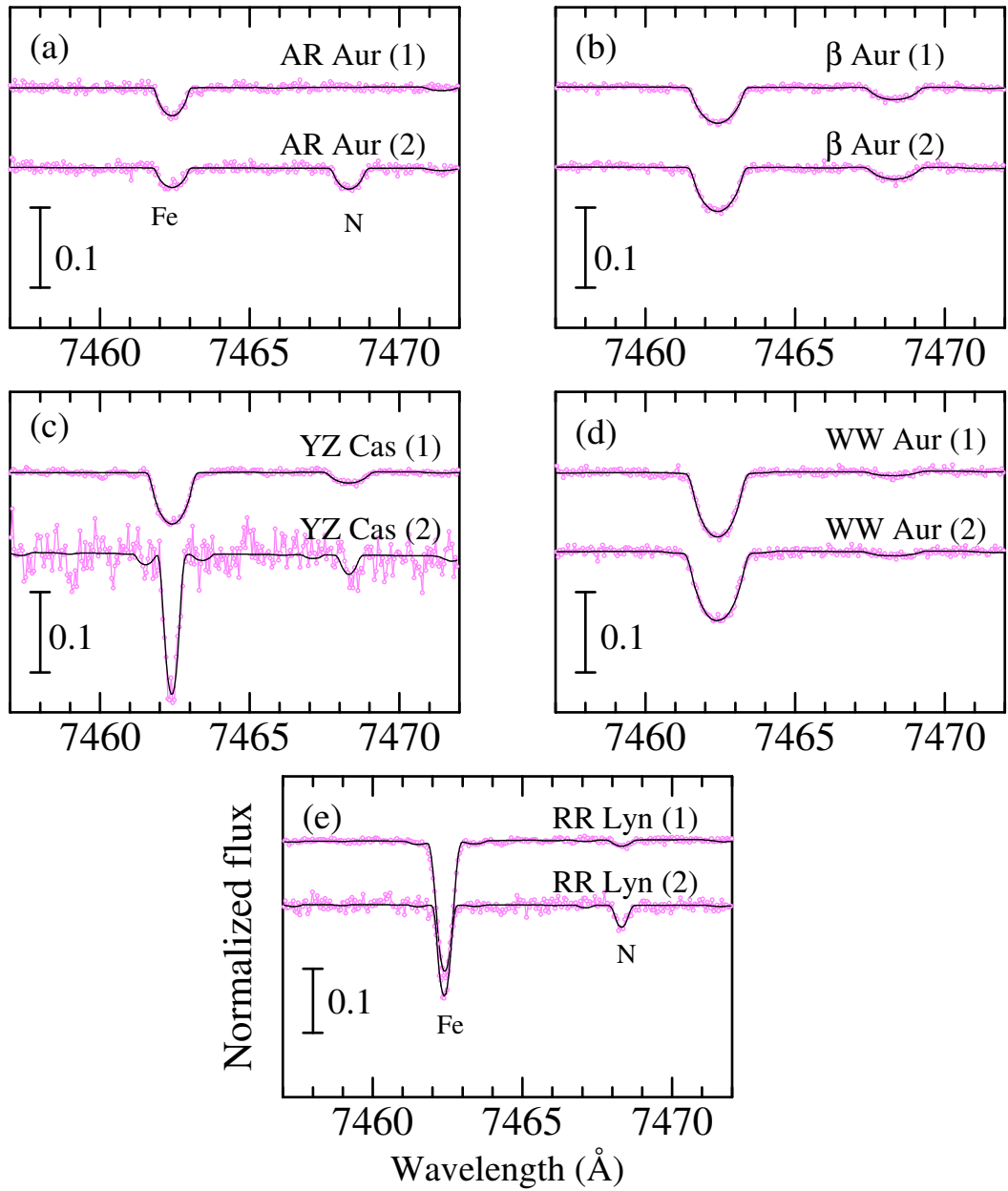}
\caption{
Synthetic spectrum fitting in the 7457--7472~\AA\ region 
comprising the N~{\sc i}~7468 line (along with the line
of Fe). Otherwise, the same as in Fig.~4.
}
\label{fig5}
\end{minipage}
\end{figure*}

\setcounter{figure}{5}
\begin{figure*}
\begin{minipage}{140mm}
\includegraphics[width=14.0cm]{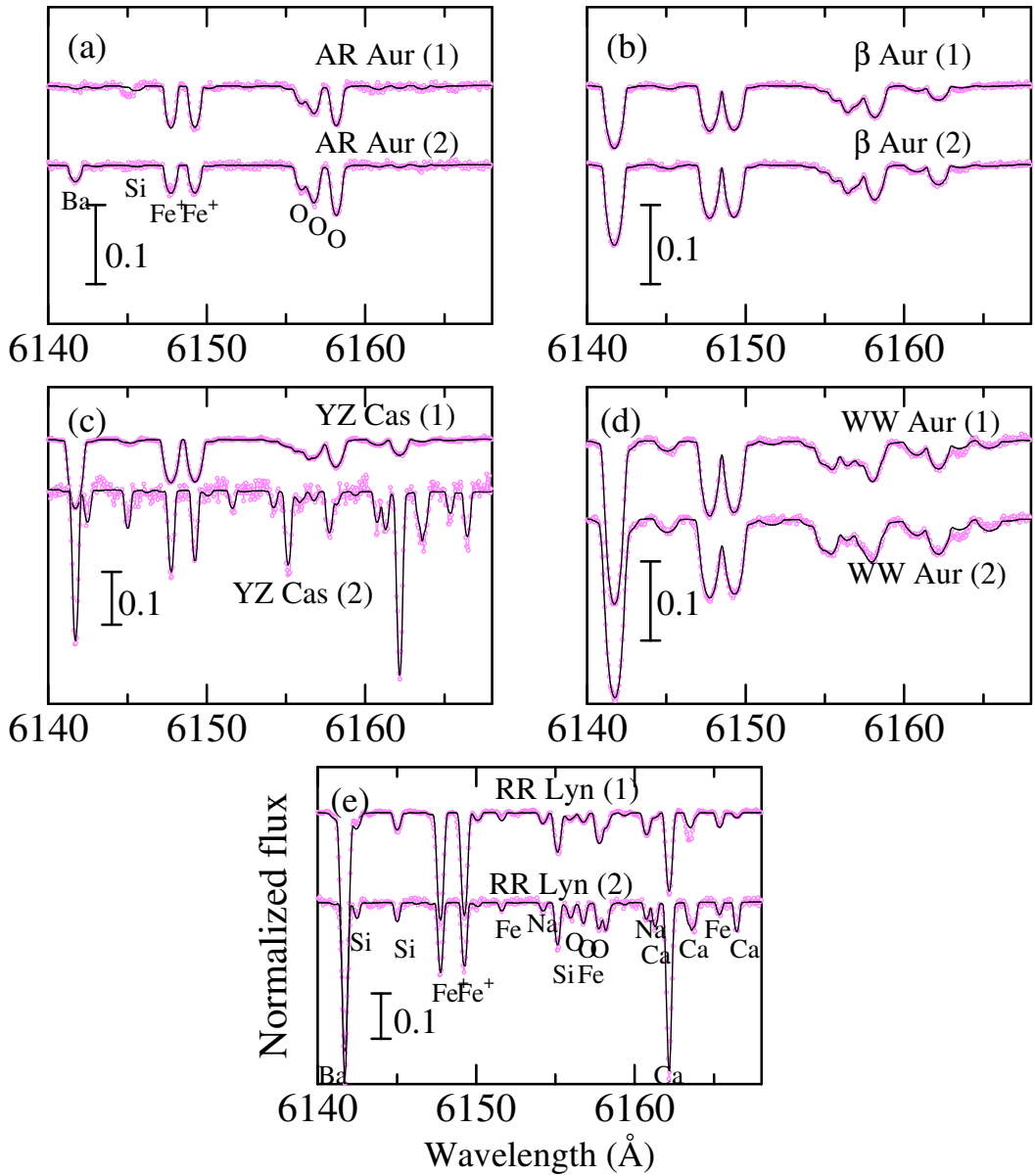}
\caption{
Synthetic spectrum fitting in the 6140--6168~\AA\ region 
comprising the O~{\sc i}~6156--8 lines (along with the lines
of Na, Si, Ca, Fe, and Ba). Otherwise, the same as in Fig.~4.
}
\label{fig6}
\end{minipage}
\end{figure*}

\setcounter{figure}{6}
\begin{figure*}
\begin{minipage}{140mm}
\includegraphics[width=14.0cm]{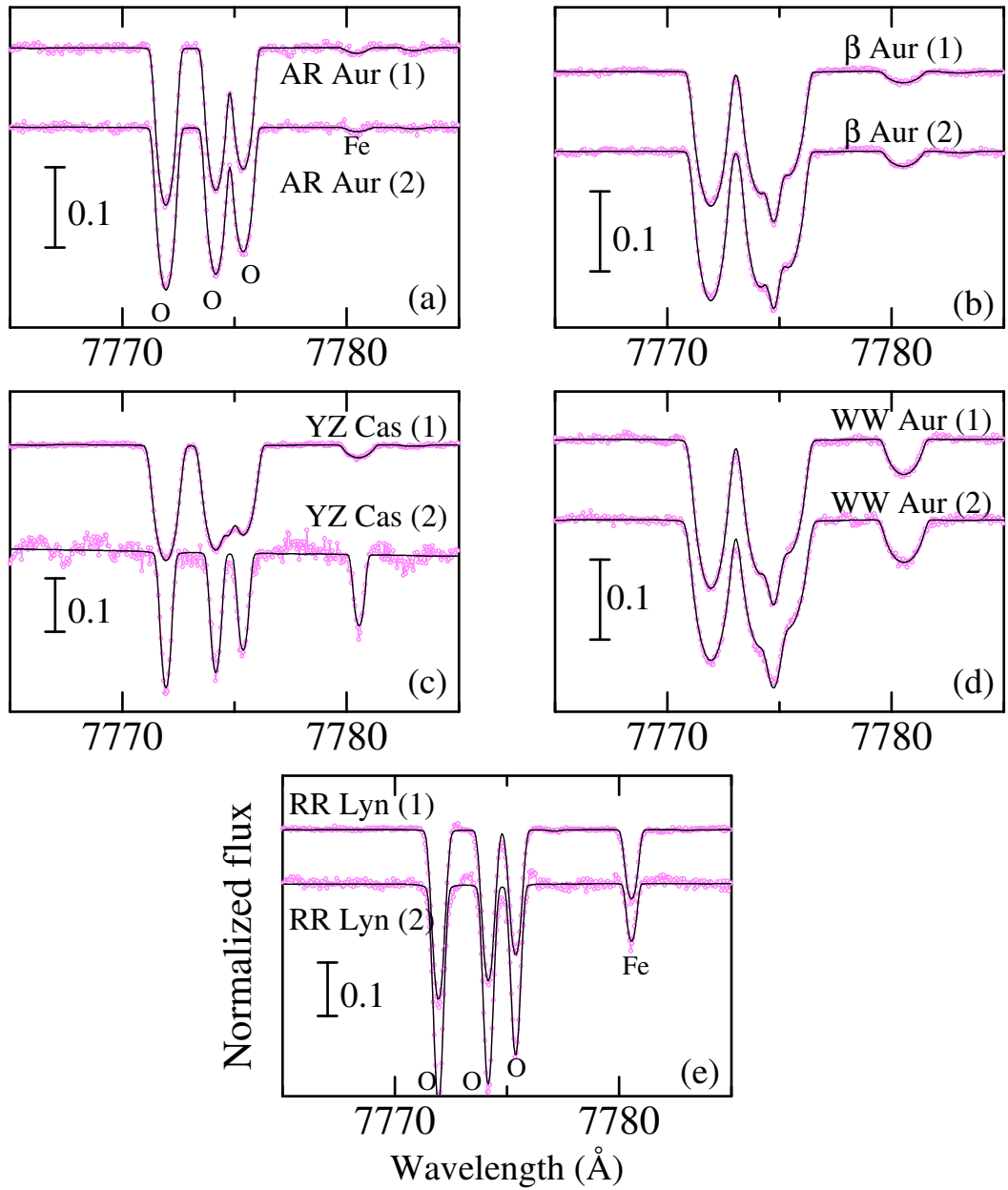}
\caption{
Synthetic spectrum fitting in the 7765--7785~\AA\ region 
comprising the O~{\sc i}~7771--5 lines (along with the line
of Fe). Otherwise, the same as in Fig.~4.
}
\label{fig7}
\end{minipage}
\end{figure*}

\setcounter{figure}{7}
\begin{figure*}
\begin{minipage}{140mm}
\begin{center}
\includegraphics[width=14.0cm]{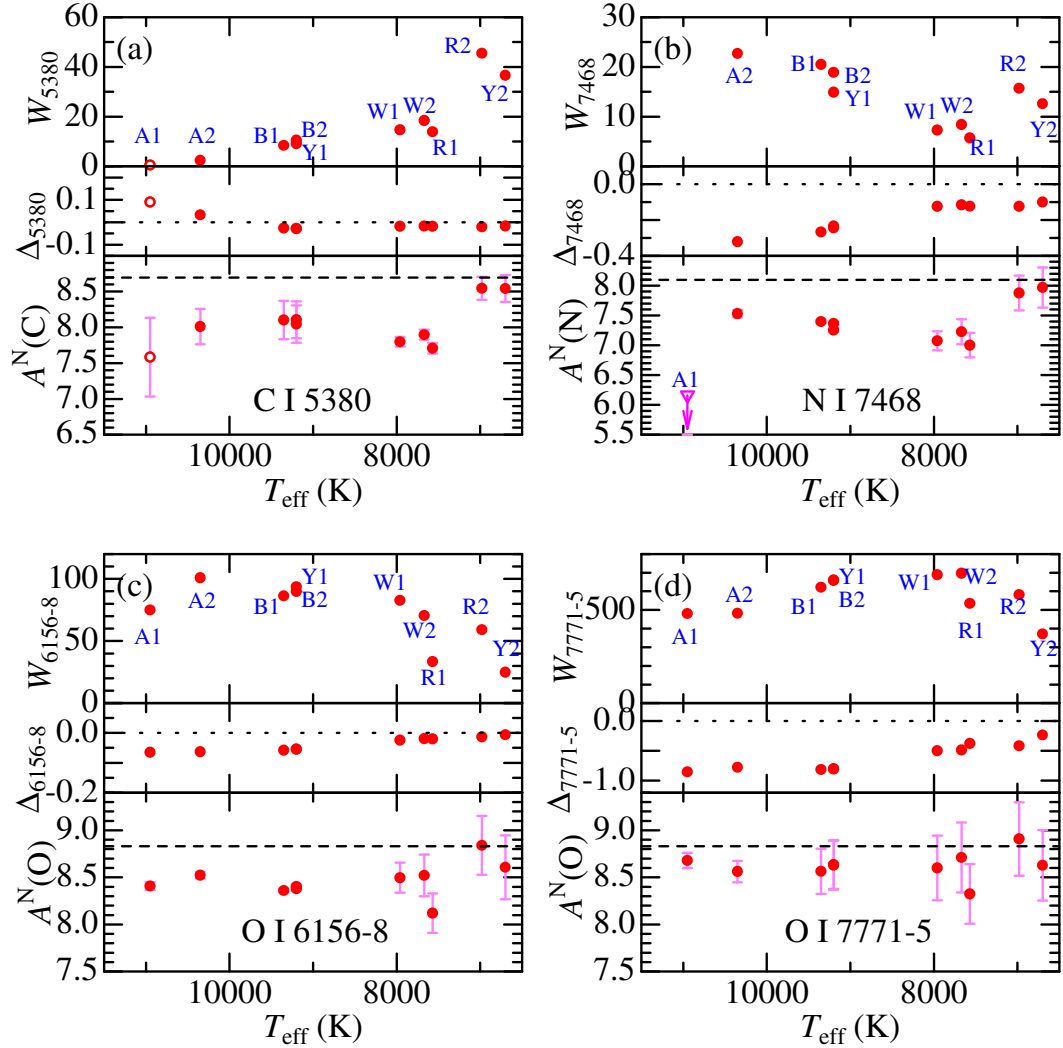}
\caption{
(a) C~{\sc i}~5380-related quantities plotted against $T_{\rm eff}$:
$W_{5380}$ (equivalent width), $\Delta_{5380}$ (non-LTE 
correction), and $A^{\rm N}$(C) (non-LTE logarithmic number abundance of C
expressed in the usual normalization of $A_{\rm H} = 12$).
The error bars in $A^{\rm N}$(C) are the combined uncertainties due to those 
of atmospheric parameters and photometric errors (cf. Sect.~4.3 in Paper~I).
The codes attached to the symbols are almost self-explanatory;
e.g., A1 is AR~Aur~(1), B2 is $\beta$~Aur~(2), or Y2 is YZ~Cas~(2). 
The reference $A^{\rm N}$(C) value for the standard star Procyon is 
shown by the horizontal dashed line.    
(b) Similar to panel (a) but for N~{\sc i}~7460.
(c) Similar to panel (a) but for O~{\sc i}~6156--8.
(d) Similar to panel (a) but for O~{\sc i}~7771--5.
Note that the open symbols in panel (a) indicate the results 
with considerably large uncertainties and the downward arrow
in panel (b) means the upper limit, which correspond to  
C and N abundances for AR~Aur~(1), respectively.
}
\label{fig8}
\end{center}
\end{minipage}
\end{figure*}

\setcounter{figure}{8}
\begin{figure*}
\begin{minipage}{140mm}
\begin{center}
\includegraphics[width=14.0cm]{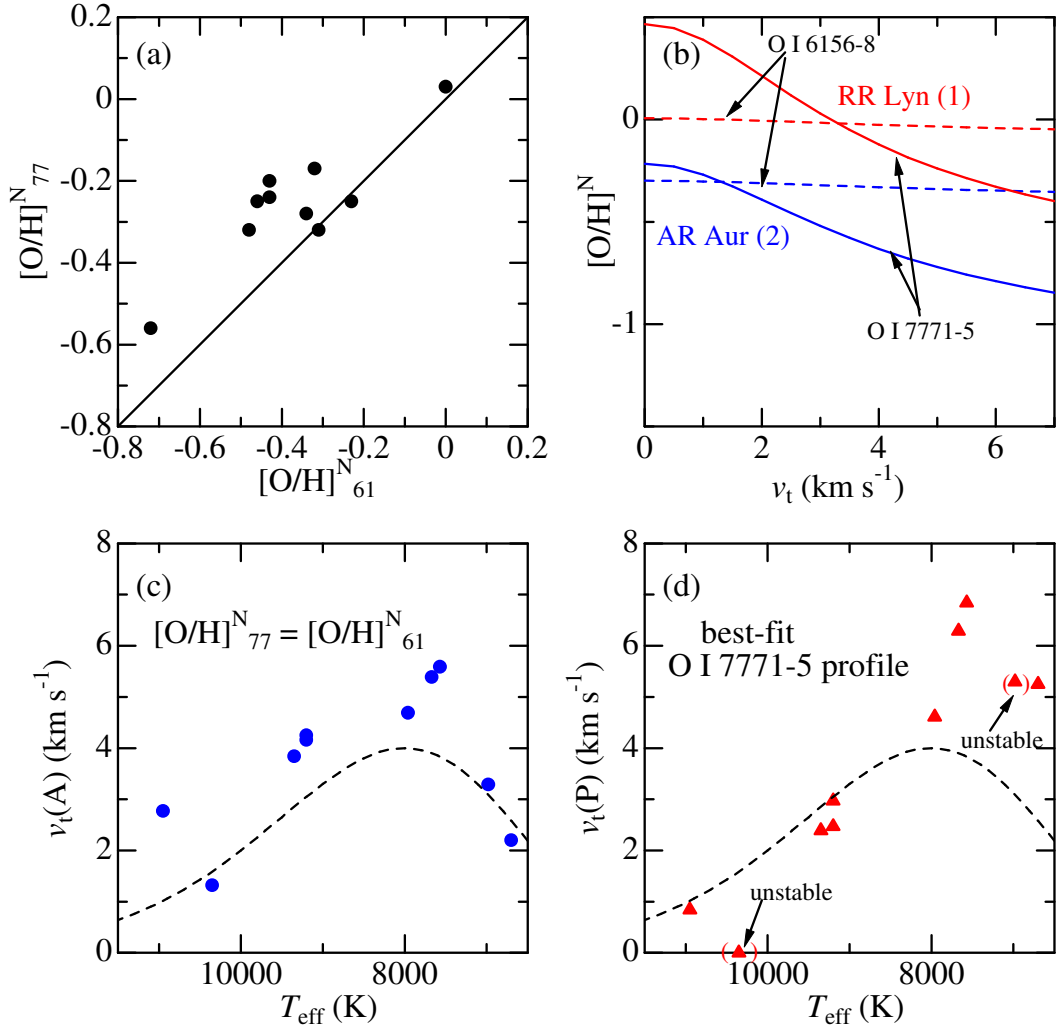}
\caption{(a) Comparison of two kinds of non-LTE oxygen abundances
(relative to Procyon) derived from O~{\sc i} 7771--5 and 
O~{\sc i} 6156--8 lines. (b) Examples of how $v_{\rm t}$(A) 
(microturbulence derived by requiring the abundance consistency
between [O/H]$^{\rm N}_{77}$ and [O/H]$^{\rm N}_{61}$) 
can be determined, shown for the cases of AR~Aur~(2) and RR~Lyn~(1). 
(c) Such determined ``abundance-matched'' $v_{\rm t}$(A) results plotted 
against $T_{\rm eff}$.
(d) The values of $v_{\rm t}$(P) (``profile-based'' microturbulence derived 
from the profile-fitting of O~{\sc i} 7771-5 lines) plotted against $T_{\rm eff}$. 
Since the $v_{\rm t}$(P) solutions for AR~Aur~(2) and RR~Lyn~(2) 
did not converge showing oscillatory behaviors, only the rough
(mean) values are shown only for reference, as indicated by parentheses. 
The dashed line shown in panels (c) and (d) is the $v_{\rm t}$ vs. 
$T_{\rm eff}$ relation adopted in this study.
}
\label{fig9}
\end{center}
\end{minipage}
\end{figure*}

\setcounter{figure}{9}
\begin{figure*}
\begin{minipage}{70mm}
\begin{center}
\includegraphics[width=7.0cm]{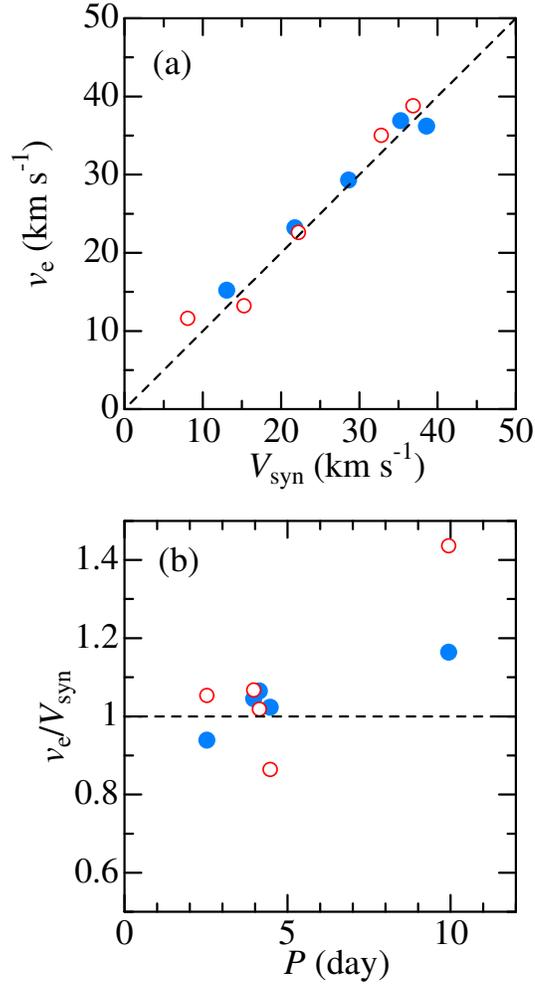}
\caption{
(a) Correlation between the observed equatorial rotational velocity 
[$v_{\rm e} \equiv (v_{\rm e}\sin i)/\sin i_{\rm orb}$: assuming the alignment 
of orbital and rotational axes] and the predicted rotational velocity 
based on the assumption of synchronization [$V_{\rm syn} \equiv 2 \pi R /P $)].
The primary and secondary components are denoted by filled and open symbols, respectively.
(b) $v_{\rm e}/V_{\rm syn}$ ratios plotted against $P$ (period of orbital motion). 
}
\label{fig10}
\end{center}
\end{minipage}
\end{figure*}

\setcounter{figure}{10}
\begin{figure*}
\begin{minipage}{140mm}
\begin{center}
\includegraphics[width=14cm]{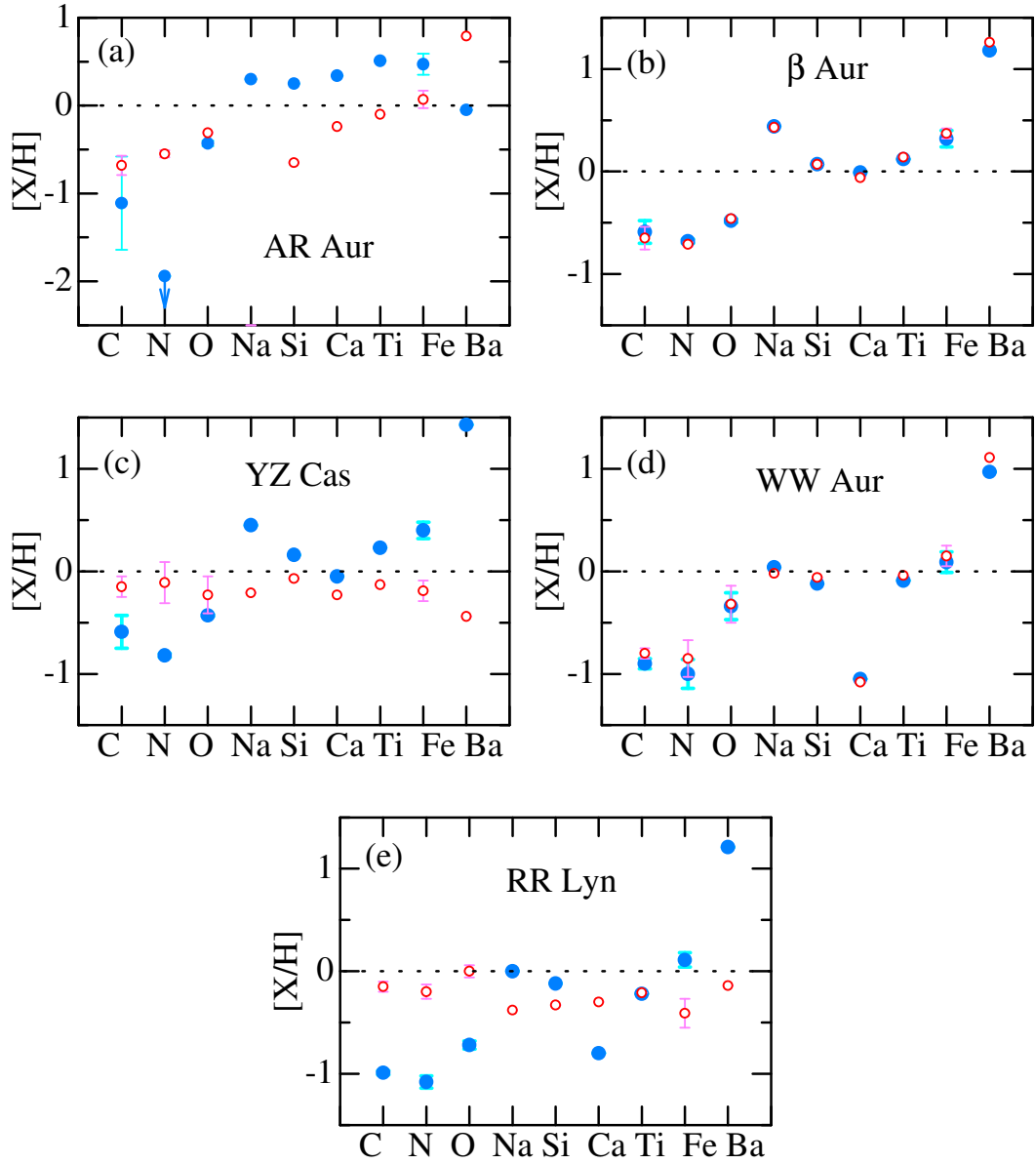}
\caption{
Graphical plots of the [X/H] values given in Table~7. The primary and secondary 
components are denoted by filled and open symbols, respectively.
The attached error bars to [C/H], [N/H] and [O/H] are the same as the case
for $A^{\rm N}$ in Fig.~8, while that of [Fe/H] is the standard deviation
of the four [Fe/H] values (cf. the caption in Table~7).
(a) AR~Aur, (b) $\beta$~Aur, (c) YZ~Cas, (d) WW~Aur, and (e) RR~Lyn. 
}
\label{fig11}
\end{center}
\end{minipage}
\end{figure*}

\setcounter{figure}{11}
\begin{figure*}
\begin{minipage}{70mm}
\begin{center}
\includegraphics[width=7cm]{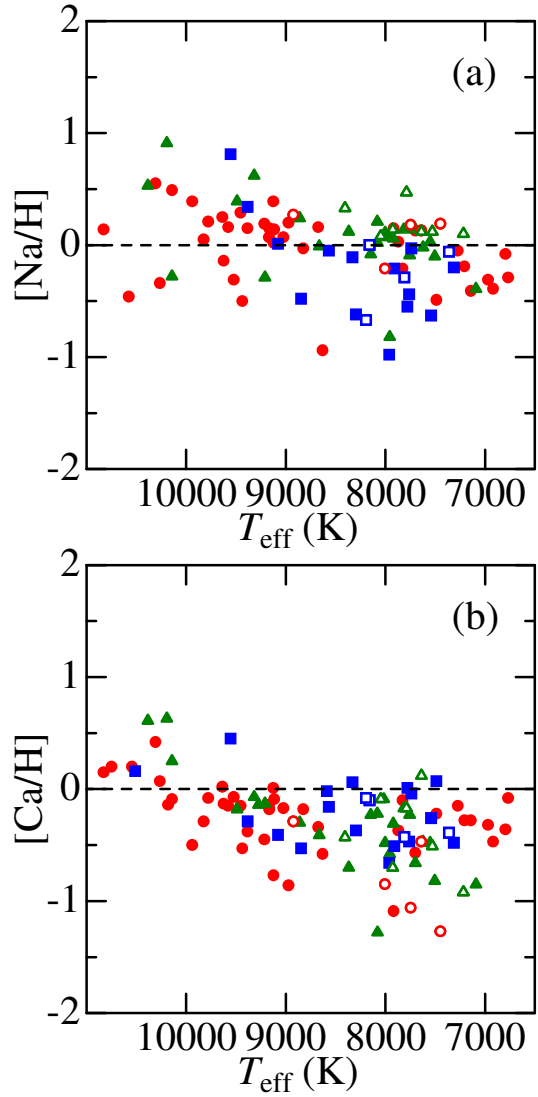}
\caption{
(a) [Na/H] vs. $T_{\rm eff}$ and (b) [Ca/H] vs. $T_{\rm eff}$ relations
for 100 A-type main-sequence stars studied in Paper~I (these abundance data
are given in ``paperI\_suppl.dat'' of the online material).
Corresponding $v_{\rm e}\sin i$ ranges are discriminated by the types of symbols:
circles ($0 < v_{\rm e}\sin i < 30$~km~s$^{-1}$), 
triangles ($30 \le v_{\rm e}\sin i < 70$~km~s$^{-1}$), 
and squares ($70 \le v_{\rm e}\sin i < 100$~km~s$^{-1}$).
Open symbols denote 16 stars belonging to the Hyades cluster.
}
\label{fig11}
\end{center}
\end{minipage}
\end{figure*}

\setcounter{figure}{12}
\begin{figure*}
\begin{minipage}{140mm}
\begin{center}
\includegraphics[width=12cm]{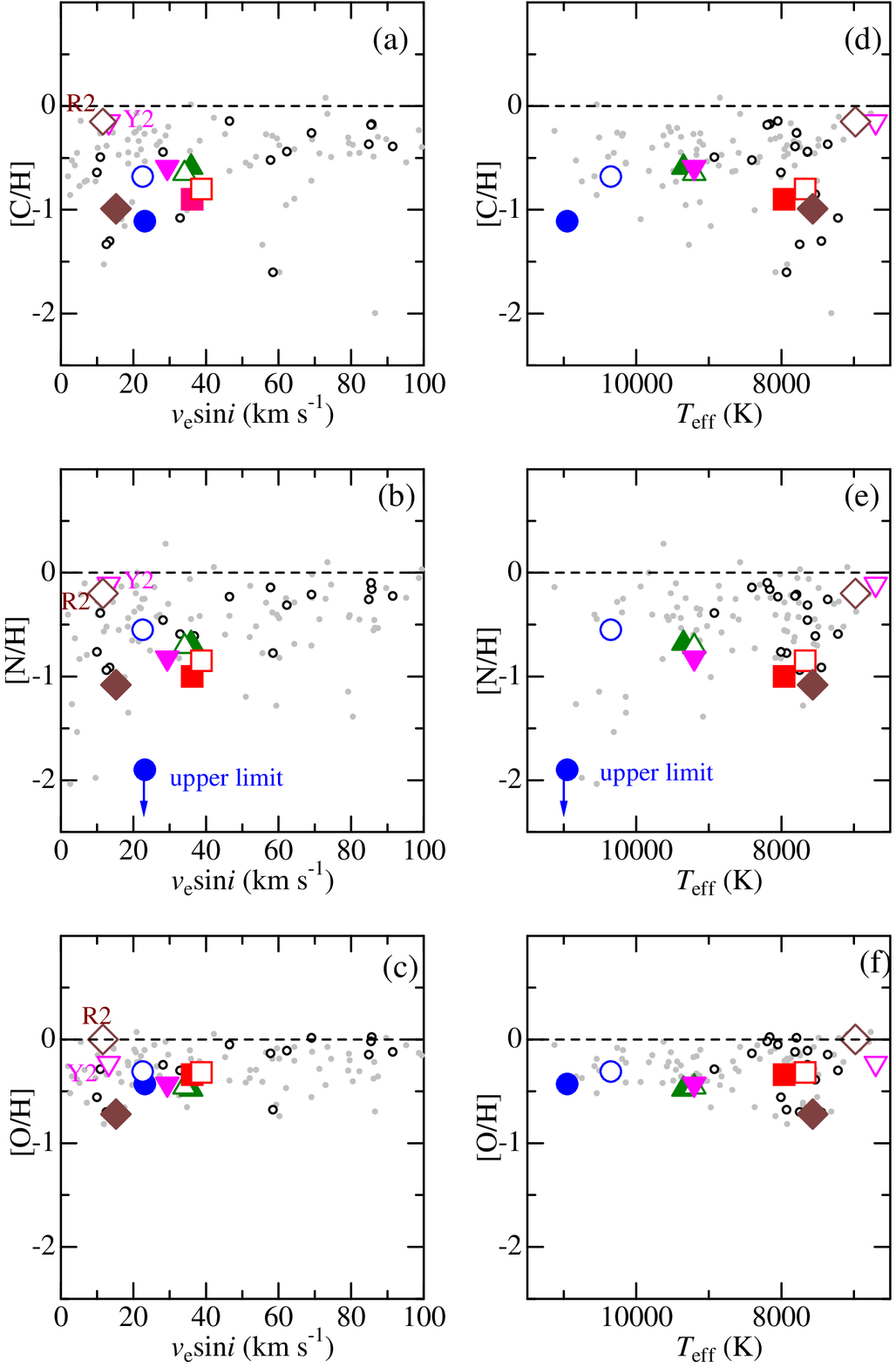}
\caption{
Comparison of the CNO abundances of the 10 program stars (large symbols; the same
meaning as in Fig.~3) with those of 100 A-type stars derived in Paper~1 (small symbols;
where Hyades stars are denoted by small open circles, while others are by gray dots). 
In the left (a--c) and right (d--f) panels are plotted the results against 
$v_{\rm e}\sin i$ and $T_{\rm eff}$, while the top (a, d), middle (b, e), and 
bottom (c, f)  panels correspond to  [C/H], [N/H], and [O/H], respectively. 
In the left panels (a)--(c), RR~Lyn (2) and YZ~Cas (2) are denoted as R2 and Y2,
in order to remark that these two are normal abundance stars.
}
\label{fig11}
\end{center}
\end{minipage}
\end{figure*}


\begin{thebibliography}{}
\bibitem[]{}
  Bressan A., Marigo P., Girardi L., Salasnich B., Dal Cero C., Rubele S., 
  Nanni A., 2012, MNRAS, 427, 127
\bibitem[]{}
  Burkhart C., \& Coupry M.~F., 1991, A\&A, 249, 205
\bibitem[]{}
  Eker Z., Bilir S., Soydugan F., Yaz G\"{o}k\c{c}e E., Soydugan E., T\"{u}ys\"{u}z M.,
  \c{S}eny\"{u}z T., Demircan O., 2014, PASA, 31, e024
\bibitem[]{}
  Folsom C.~P., Kochukhov O., Wade G.~A., Silvester J., Bagnulo S., 2010, MNRAS, 407, 2383
\bibitem[]{}
  Hui-Bon-Hoa A., 2000, A\&AS, 144, 203
\bibitem[]{}
  Iliji\'{c} S. 2004, in Spectroscopically and Spatially Resolving the Components
  of Close Binary Stars, ASP Conf. Ser. Vol. 318, ed. R. W. Hilditch, H. Hensberge,
  \& K. Pavlovski (San Francisco: Astronomical Society of the Pacific), 107 
\bibitem[]{}
  Khokhlova V.~L., Zverko Yu.,Zhizhnovskii I., Griffin R.~E.~M., 1995, Astron. Lett., 21, 818
\bibitem[]{}
  Kitamura M., Kim T.-H., Kiyokawa M., 1976, Ann. Tokyo Astron. Obs. 2nd. Ser., 16, 22
\bibitem[]{}
  Kitamura M., Kondo M., 1978, Ap\&SS, 56, 341
\bibitem[]{}
  Kondo M., 1976, Ann. Tokyo Astron. Obs. 2nd. Ser., 16, 1
\bibitem[]{}
  Kurucz R.~L., 1993, Kurucz CD-ROM, No. 13, ATLAS9 Stellar Atmosphere
  Program and 2 km/s Grid (Cambridge, MA: Harvard-Smithsonian Center
  for Astrophysics)
\bibitem[]{}
  Kurucz R.~L., Bell B., 1995, Kurucz CD-ROM, No. 23,
  Atomic Line Data (Cambridge, MA: Harvard-Smithsonian Center
  for Astrophysics)
\bibitem[]{}
  Lyubimkov L.~S., Rachkovskaya, T.~M., 1995 Astron. Rep., 39, 63
\bibitem[]{}
  Lyubimkov L.~S., Rachkovskaya T.~M., Rostopchin S.~I., 1996, Astron. Rep., 40, 802
\bibitem[]{}
  Pavlovski K., Southworth J., Kolbas V., Smalley B., 2014, MNRAS, 438, 590
\bibitem[]{}
  Pavlovski K., Southworth J., Tamajo E., 2008, Contrib. Astron. Obs. Skalnat\'{e} Pleso, 38, 437
\bibitem[]{}
  Popper D.~M., 1971, ApJ, 169, 549
\bibitem[]{}
  Preston G.~W., 1974, ARA\&A, 12, 257
\bibitem[]{}
  Richer J., Michaud G., Turcotte S., 2000, ApJ, 529, 338
\bibitem[]{}
  Samus N.~N., Kazarovets E.~V., Durlevich O.~V., Kireeva N.~N., Pastukhova E.~N., 2017,
  Astron. Rep., 61, 80
\bibitem[]{}
  Southworth J., Smalley B., Maxted P.~F.~L., Claret A., Etzel P.~B., 2005, MNRAS, 363, 529
\bibitem[]{}
  Southworth J., Bruntt H., Buzasi D.~L., 2007, A\&A, 467, 1215
\bibitem[]{}
  Takeda Y., Han I., Kang D.-I., Lee B.-C., Kim K.-M., 2008,
  JKAS, 41, 83
\bibitem[]{}
  Takeda Y., Hashimoto O., Honda S., 2018b, ApJ, 862, 57
\bibitem[]{}
  Takeda Y., Jeong G., Han I., 2018a, PASJ, 70, 8
\bibitem[]{}
  Takeda Y., Kawanomoto S., Ohishi N., Kang D.-I., 
  Lee B.-C., Kim K.-M., Han I., 2018c, PASJ, 70, 91 (Paper~I)
\bibitem[]{}
  Takeda Y., Sadakane K., 1997, PASJ, 49, 571
\bibitem[]{}
  Tomkin J., Fekel F. C., 2006, AJ, 131, 2652
\bibitem[]{}
  Torres G., Lacy C.~H.~S., Pavlovski K., Fekel F.~C., Muterspaugh M.~W., 
  2015, ApJ, 150, 154 
\bibitem[]{}
  Varenne O., Monier R., 1999, A\&A, 351, 247
\bibitem[]{}
  Wilson R.~E., Van Hamme W., 2014, ApJ, 780, 151
\end{thebibliography}
\end{document}